\documentclass[twocolumn]{aastex62}

\usepackage{braket}
\usepackage{mathrsfs}

\begin{document}
\title{Abundances of ordinary chondrites in thermally evolving planetesimals}

\email{shigeru@elsi.jp}
\author[0000-0002-3161-3454]{Shigeru Wakita}
\affiliation{Center for Computational Astrophysics, National Astronomical Observatory of Japan, Mitaka, Tokyo 181-8588, Japan}
\affiliation{Earth-Life Science Institute, Tokyo Institute of Technology, Meguro-ku, Tokyo 152-8550, Japan}
\author{Yasuhiro Hasegawa}
\affiliation{Jet Propulsion Laboratory, California Institute of Technology, Pasadena, CA 91109, USA}
\author{Takaya Nozawa}
\affiliation{Division of Theoretical Astronomy, National Astronomical Observatory of Japan, Mitaka, Tokyo 181-8588, Japan}

\begin{abstract}
Chondrites are one of the most primitive objects in the solar system, and keep the record of the degree of thermal metamorphism experienced in their parent bodies.
This thermal history can be classified by the petrologic type.
We investigate the thermal evolution of planetesimals to account for the current abundances (known as the fall statistics) of petrologic types 3 - 6 ordinary chondrites.
We carry out a number of numerical calculations in which formation times and sizes of planetesimals are taken as parameters.
We find that planetesimals that form within 2.0 Myr after the formation of Ca-Al-rich inclusions (CAIs) can contain all petrologic types of ordinary chondrites. 
Our results also indicate that plausible scenarios of planetesimal formation, which are consistent with the fall statistics, 
are that planetesimals with radii larger than 60 km start to form around 2.0 Myr after CAIs and/or that ones with radii less than 50 km should be formed within 1.5 Myr after CAIs.
Thus, thermal modelling of planetesimals is important for revealing the occurrence and amount of metamorphosed chondrites, and for providing invaluable insights into planetesimal formation.
\end{abstract}


\section{Introduction} \label{intro}

Planetesimals are one of the central objects that play crucial roles in planet formation \citep[e.g.,][]{phb96,il04,mab09}.
It has been suggested for a long time that planetary cores were built from planetesimal collisions and resulting mergers in protoplanetary disks
\citep[e.g.,][]{ws89,ki98,msm10,rkm14}.
The presence of planetesimals in planet-forming regions can also affect the orbital evolution of protoplanets there,
which is referred to as planetesimal-driven migration 
\citep[e.g.,][]{ltd10,kdm16}.
More recently, planetesimals has been regarded as the main contributor for metal enrichment of Jovian planets both in the solar system and extrasolar planetary systems 
\citep[e.g.,][]{sg04,mf11,mbm16}.
These are all regulated by the dynamics of planetesimals,
and hence are determined by the properties of planetesimals such as their abundances and masses.
Thus, it is of fundamental importance to identify and characterize when and how planetesimal formation takes place in protoplanetary disks 
for fully exploring the formation and migration histories of planets.

Understanding of planetesimal formation is still far from complete despite recent progress \citep[e.g.,][]{jbmt14}.
There are two reasons for this. One is that, theoretically, no force (i.e., surface adhesion force or gravity) is effective for km-sized bodies.
Many of theoretical studies thereby attempt to establish high density regions of solids in protoplanetary disks 
that eventually become gravitationally unstable and lead to direct and instant formation of planetesimals due to self-gravity
\citep[e.g.,][]{chp01,jom07,kl07}.
The other is that, observationally, planetesimals are invisible at almost all wavelengths in the nearby star-forming regions.
This arises from the fact that emission from gas and dust and their scattered light become generally dominant in the observations.
Investigating planetesimals is therefore not straightforward even for nearby, young stars.

The solar system stands out in exploring planetesimal formation.
This is because we can obtain key information from asteroids and meteorites.
It is reasonable to consider that asteroids are the remnants of planetesimals 
that survived a number of constructive and destructive events occurring over the age of the solar system 
\citep[e.g.,][]{bdn05,bdn05b,mbn09,zlk17,tmd18}.
In other words, asteroids are unique objects that contain a wealth of information about planetesimals.
There are completed and ongoing sample return missions for investigating the composition of asteroids: Hayabusa, OSIRIS-REx, and Hayabusa 2.
For instance, the Hayabusa revealed that the chemical and isotopic compositions of particles on the asteroid Itokawa are similar to those of meteorites falling onto the Earth \citep[e.g.,][]{nnt11,ykf15}.
Asteroids can also be used as a probe to specify the size distribution of planetesimals.
Indeed, theoretical studies suggest that primordial planetesimals would be larger than 100 km in diameter in order to explain the size distribution of current main belt asteroids \citep[e.g.,][]{bdn05,mbn09,tmd18}.
Thus, asteroids give a number of invaluable insights about planetesimals.
It is however important to realize that these insights are derived from the current status of asteroids, and hence one has to take into account the full history of a number of events that occurred for asteroids.

Meteorites act as important and complimentary objects for understanding planetesimal formation \citep[e.g.,][]{m99}.
This is because they literally contain the fossil record of how the solar system formed.
For example, chondrites are known as the most primitive meteorites and contain the first condensed materials in the solar system, called Ca-Al-rich inclusions (CAIs) \citep[e.g.,][]{m05}.
It is well known that the age of the solar system is measured from CAIs, which is 4567.30$\pm$0.16 million years \citep[e.g.,][]{mdz95,aki10,cbk12,dm14}.
Also, the detailed chemical and isotopic analyses of meteorites suggest that planetesimals should have formed soon after CAI formation and continued for about a few million years \citep{dac14,gtb14,bac17}.
Moreover, meteorites contain another primitive component called chondrules that are considered to have formed within 5 Myr after CAI formation \citep[e.g.,][]{cbk12, bcb15}.
The formation mechanisms of chondrules are currently under active investigation, one of which is highly relevant to planetesimal collisions \citep[e.g.,][]{jmm15,hwmo16,htm16,wmoh17,mohw17}.

Another remarkable feature of meteorites is that more than forty-thousand of them are currently found.
Interestingly, 90\% of the analyzed meteorites are ordinary chondrites \citep[Meteoritical Bulletin website \url{http://www.lpi.usra.edu/meteor/metbull.php}; ][]{gpm14}.
Thanks to such abundant samples, statistical studies become possible, including classification.
In general, meteorites are grouped by their chemical composition and texture \citep[e.g.,][]{vw67,kkgs05,sk05,wmk06}.
We here focus on the petrologic type that links to the degree of thermal metamorphism.
The degree of metamorphism is considered to reflect the maximum temperatures that the parent bodies of meteorites experienced \citep[e.g.,][]{d81,sk05,hrg06}. 
Type 3 chondrites are the least metamorphosed ones and would not pass through temperatures higher than $\sim$600 $^\circ$C. 
Type 6 chondrites are the most highly metamorphosed, possibly in the environment above $\sim$800 $^\circ$C, and type 4 and 5 chondrites would be between them \citep{k00,w05,sk05,hrg06}.
Since the experienced temperatures depend on the position inside planetesimals, previous studies make use of thermal evolution models of planetesimals \citep[e.g.,][]{mgg02,hgt13,gtb14}.
Attempts to reproduce the fall statistics of ordinary chondrites \citep{gm00,mgg02} have not worked well under the assumption that parent bodies of meteorites are constituted of single sized planetesimals.

It is, however,  highly possible that ordinary chondrites originate from a variety of planetesimals \citep[e.g.,][]{mtb13,vzn15,bac17}.
Furthermore, the internal temperature of planetesimals is sensitive not only to the location within the planetesimals, but also to their formation times and initial sizes \citep[e.g.,][]{gtb14,wni14,lgg16,rba17}.
In this paper, therefore, we extensively investigate the thermal evolution of planetesimals with a wide range of model parameters such as formation times and sizes of planetesimals.
By comparing our numerical results with the fall statistics of ordinary chondrites in each petrologic type, we will specify the most plausible values of the formation times and size distributions of planetesimals.

The plan of this paper is as follows.
In Section \ref{sec:mod}, we describe how we calculate the thermal evolution of planetesimals with model parameters.
Our numerical results are shown in Section \ref{sec:res}, where we discuss the dependencies of the model parameters on the results.
In Section \ref{sec:disc}, we compare our results with the fall statistics of ordinary chondrites and discuss implications for planetesimals formation.
Our conclusions are given in Section \ref{sec:conc}.

\section{Methods}\label{sec:mod}

\subsection{Thermal evolution models of planetesimals} \label{mod:thermal}

We adopt thermal evolution models of planetesimals that were constructed in our previous work \citep{wni14,wnh17}. 
We assume that planetesimals are spherical and have the spatially uniform chemical composition.
It is also assumed that, once planetesimals form with a radius of $R_{\rm pl}$ at a given time ($t=t_{\rm pl}$), they do not experience any further growth and destructive processes afterward.
Then, thermal evolution of a planetesimal is numerically calculated by solving the equation of heat conduction,
\begin{equation}
 \rho c \frac{\partial T(r,t)}{\partial t} = \frac{1}{r^2}\frac{\partial}{\partial r} \left( r^2 K \frac{\partial T(r,t)}{\partial r} \right) + A \exp(-\lambda t) \label{eq:heat},  
\end{equation}
where $t$ is the time measured since the formation of the planetesimal, $r$ is the distance from the center of the planetesimal, 
$T(r, t)$ is the temperature of materials located at $r$ in the planetesimal at a time $t$,
$A$ is the radiogenic heat generation rate per unit volume, and $\lambda$ is the decay constant of the radionuclides.
In our simulations, decay heat arising from the short-lived radionuclide, $^{26}$Al is taken as the heating source of planetesimals \citep[e.g.,][]{mft82}.
The initial temperature of planetesimals is set at $T(r,t=t_{\rm pl}) = -123 ^\circ$C (150 K), which is low enough not to affect the results of our calculations.
We assume that physical quantities, such as $\rho$, $c$, and $K$, are constant for simplicity; 
$\rho$ = 3300 kg m$^{-3}$, $c$ = 910 J kg$^{-1}$ K$^{-1}$, and $K$ = 2 J s$^{-1}$ m$^{-1}$ K$^{-1}$ are the adopted values of the bulk density, specific heat, and thermal conductivity, respectively \citep{ym83,ocb10}.

If planetesimals contain volatiles such as water, CO$_2$ and organics, then different values should be adopted for the above three quantities. 
Furthermore, if this would be the case, the temperature distribution of planetesimals would be regulated not only by thermal conductivity, 
but also by transport of such volatiles. 
However, these complexities should be taken into account in exploring thermal evolution of volatile-rich bodies 
such as parent bodies of carbonaceous chondrites \citep[e.g.,][]{gm89,ts05,bt17}. 
In this paper, we focus on parent bodies of ordinary chondrites that accreted in very volatile-poor environments, and most of them are (highly) thermally metamorphosed.
Therefore, the impact of volatiles can be negligible, except for the ones which contain a small amount of volatiles and/or experience the least metamorphism.
In this sense, the physical quantities used here are reasonable, and considering heat conduction is valid for thermal evolution of planetesimals in the context of this paper.

\subsection{Key model parameters}\label{sec:para}

\begin{deluxetable*}{lcc}
\tablecaption{Summary of parameters in this study \label{table:params}}
\tablehead{Parameters of planetesimals & Symbol & Value }
\startdata
Radius (km)                                           & $R_{\rm pl}$     & 1 - 500                              \\
Power-law index for size distribution            & $\alpha$            & 0.0, 2.8, 3.5, 4.5                \\
Initial time of formation (Myr)                 & $t_{\rm int}$      & 0.1, 1.0, 1.5, 2.0, 2.5, 3.0  \\
Final time of formation  (Myr)                 & $t_{\rm fin}$      &  $t_{\rm int}$, 5.0 or 7.0 \\
Timescale for the formation rate (Myr)   & $\tau_f $           & 1.0, 4.0 \tablenotemark{a} \\
\enddata 
\tablenotetext{a}{The formation rate of planetesimals is assumed to be constant ($\tau_f=\infty$) for all of the simulations, except for in Section \ref{res:rate}.} 
\end{deluxetable*}

The main purpose of this study is to carry out a parameter study to determine the fundamental quantities of planetesimals.
In this paper, we introduce five parameters (see Table \ref{table:params}):
the radius of planetesimals ($R_{\rm pl}$), the power-law index ($\alpha$) for the size distribution of planetesimals, 
the initial time ($t_{\rm int}$) at which planetesimal formation begins, the final time ($t_{\rm fin}$) when planetesimal formation ends, and the formation timescale ($\tau_{f}$) of planetesimals.
In the following, we describe what values are considered for these parameters.

The typical size of planetesimals and their size distribution are hardly constrained.
Accordingly, we take the broad range from $R_{\rm pl}=$ 1 km to $R_{\rm pl}=$ 500 km.
We assume that the size distribution can be described as a power law in terms of their radius.
Then, the number ($n_{\rm pl}$) of planetesimals between $R_{\rm pl}$ and $R_{\rm pl} + dR_{\rm pl}$ can be written as
\begin{equation}
  n_{pl} = n_0 \left( R_{\rm pl}/R_{\rm pl,0} \right)^{-\alpha},
\end{equation}
for $R_{\rm pl, min} \leq R_{\rm pl} \leq R_{\rm pl, max}$, where $n_0$ is the normalization constant, and we set $R_{\rm pl, 0} = R_{\rm pl, min}$.
For the power-law index ($\alpha$), we adopt four values: $\alpha$ = 0, 2.8, 3.5, and 4.5.
In our fiducial model, we choose $\alpha$ = 2.8, following \citet{jmb15} and \citet{say16}, who propose that it would represent the primordial distribution of planetesimals.

The formation time of planetesimals is also uncertain since their formation mechanisms are still unclear.
It nonetheless might be reasonable to consider that planetesimal formation would take place when the (gaseous) solar nebula was present.
Then, we assume that formation time ($t_{\rm pl}$) is in the range of $0.1$ Myr $\leq t_{\rm pl} \leq$ 7.0 Myr.
In this parameter study, we consider that the onset time ($t_{\rm int}$) of planetesimal formation is between 0.1 Myr and 3.0 Myr.
For the termination time ($t_{\rm fin}$) of planetesimals, we consider two cases.
For one case, we assume $t_{\rm pl} = t_{\rm int}= t_{\rm fin}$ to focus on the single generation of planetesimals (see Sections \ref{res:single} and \ref{res:single2}).
For the other case, in which multiple generations of planetesimals are examined, $t_{\rm fin} = 5.0$ Myr or 7.0 Myr (see Sections \ref{res:muti1} and \ref{res:size}).
Note that in the following sections, the time is all measured since the formation of CAIs: 
CAIs are regarded as the first condensates in the solar system that formed 4567 Myr ago with the initial ratio of $^{26}$Al/$^{27}$Al $=5.25 \times 10^{-5}$ \citep[e.g.,][]{cbk12}.
Given that the abundance of $^{26}$Al decreases exponentially with a half-life of 0.72 Myr, the earlier formed planetesimals have a higher abundance of the heating source.

The formation rate of planetesimals is another important quantity that can affect global thermal histories of forming planetesimals.
In this paper, we simply assume that the formation rate ($\phi$) of planetesimals is constant with time for most of cases. 
We will also consider the case that the formation rate of planetesimals decreases exponentially with time.
Namely, it is given as $\phi = \phi_0 \exp(- t / \tau_f)$, where $\phi_0$ is the hypothetical formation rate of planetesimals in number at $t$ = 0, and $\tau_f$ is the formation timescale (see Section \ref{res:rate}).

\subsection{Normalization factors for mass fractions}\label{mod:size}

As discussed below, we compute and examine mass fractions of planetesimals that are normalized by the total mass of planetesimals under consideration.
Here, we summarize the total mass of planetesimals for three cases.

(i) The simplest case considered in this paper is that planetesimals form at a single epoch ($t_{\rm pl} = t_{\rm int} = t_{\rm fin}$) and have a single size ($R_{\rm pl}$).
In this case, the total mass of planetesimals ($M_1$) is given as
\begin{equation}
\label{eq:norm_1}
M_1 =  N_1\frac{4 \pi \rho}{3} R_{\rm pl}^3,
\end{equation}
where $N_1$ is the number of planetesimals formed.

(ii) A more realistic situation is that planetesimals form during some period ($t_{\rm int} \leq t_{\rm pl} \leq t_{\rm fin}$), but all of them have the same size ($R_{\rm pl}$).
In this case, the total mass of planetesimals ($M_2(t)$) is written as
\begin{equation}
\label{eq:norm_2}
M_2(t) =  \sum^{t}_{ t^{\prime} = t_{\rm int}} \phi (t^\prime) \frac{4 \pi \rho}{3} R_{\rm pl}^3 .
\end{equation}
When $t > t_{\rm fin}$, the summation is carried out up to $t = t_{\rm fin}$.

(iii) The most realistic case in this paper is that planetesimals form over certain times ($t_{\rm int} \leq t_{\rm pl} \leq t_{\rm fin}$), and they have a size distribution. 
Here, we assume that the power-law index ($\alpha$) and its size range ($R_{\rm pl, min}$ and $R_{\rm pl, max}$) do not change with time.
Then, the total mass of planetesimal ($M_3(t)$) is given as
\begin{equation}
M_3(t) =  \sum^{t}_{t^{\prime} = t_{\rm int}} \phi (t^\prime)  \int_{R_{\rm pl, min}}^{R_{\rm pl, max}} dR_{\rm pl} n_{\rm pl} (R_{\rm pl}) \frac{4 \pi \rho}{3} R_{\rm pl}^3 .
\label{eq:norm_3}
\end{equation}

In the next section, we present the results for these three cases.

\section{Results} \label{sec:res}

\subsection{Thermal evolution of a single-sized planetesimal formed at a single epoch}  \label{res:single}

\begin{figure*}
\plottwo{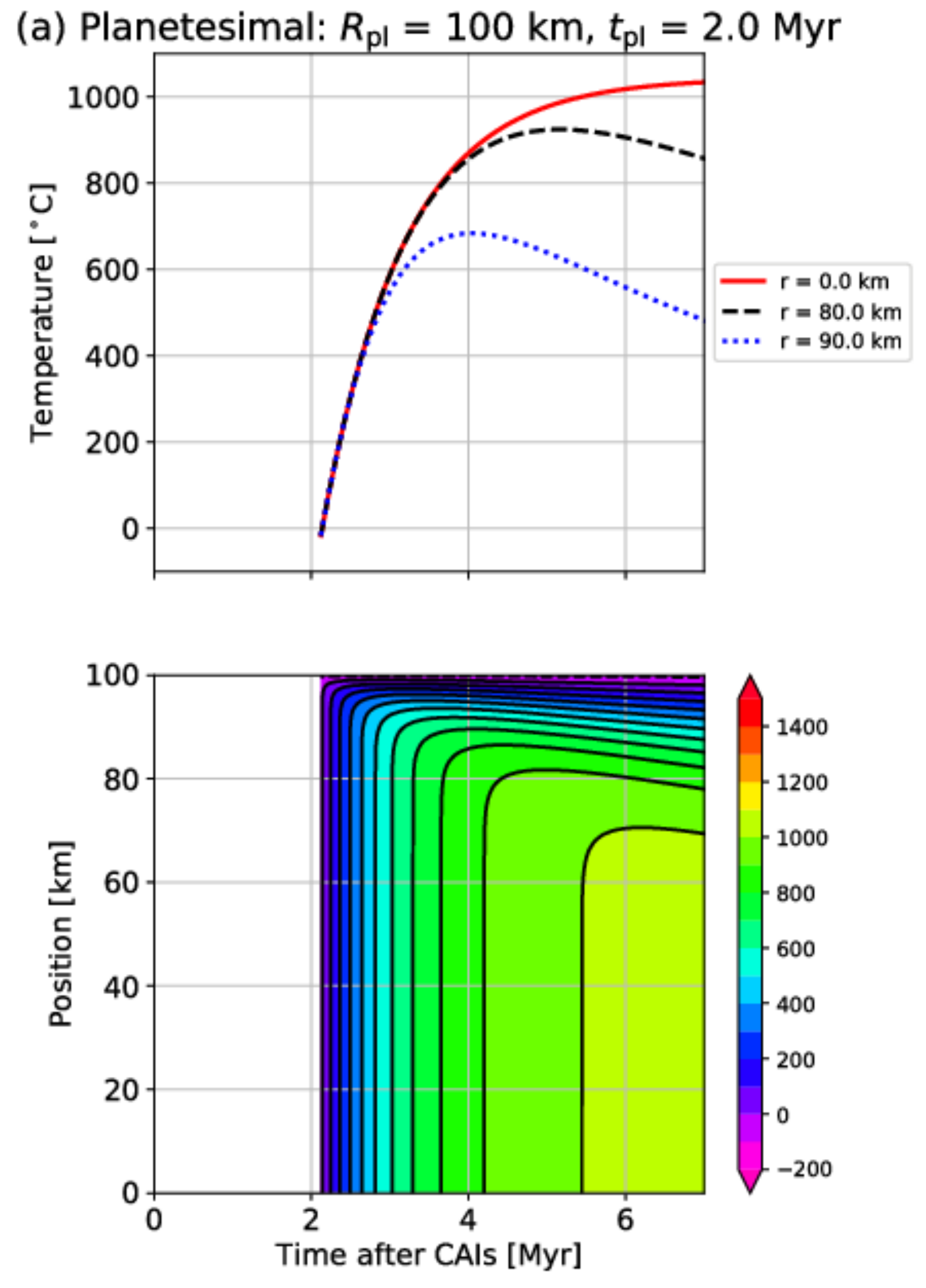}{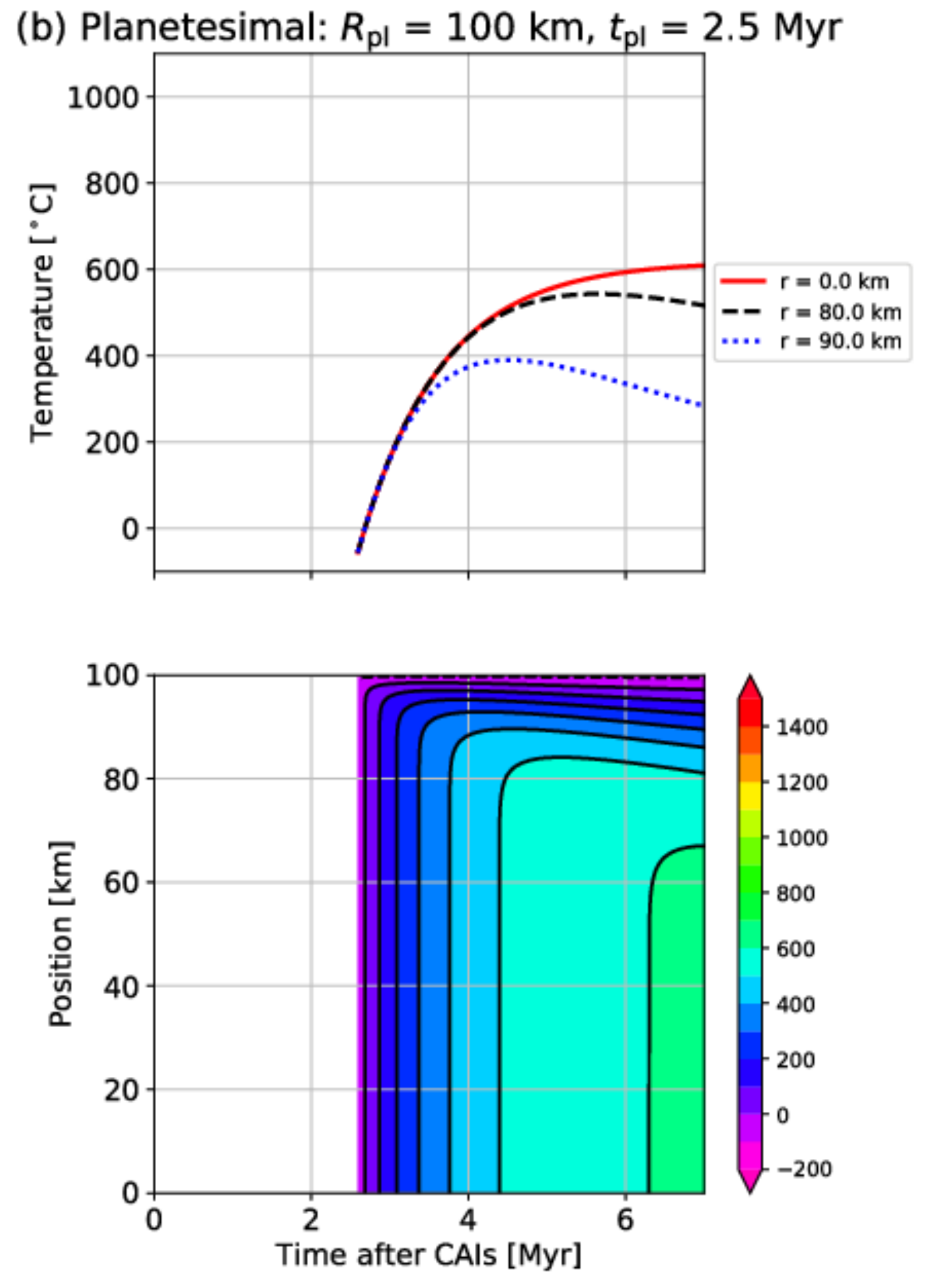}
\caption{Temperature evolutions of single-sized planetesimals ($R_{\rm pl}$ = 100 km) formed at (a) $t_{\rm pl}$ = 2.0 Myr and (b) $t_{\rm pl}$ = 2.5 Myr. 
Top panels show temperature evolutions at positions of $r$ = 0 km (red solid lines), $r$ = 80 km (black dashed lines), and $r$ = 90 km (blue dotted lines), respectively.
Bottom panels show temperatures at each position of the planetesimal. Each black line is drawn every 100 $^\circ$C.
\label{fig:single}}
\end{figure*}

In this section, we demonstrate the fundamental properties of thermal evolution of a single-sized planetesimal.
The results are obtained, using the following set of parameters: 
$R_{\rm pl} =100$ km, and $t_{\rm pl} = t_{\rm int} = t_{\rm fin} = 2.0$ Myr or 2.5 Myr, that is, a single generation of planetesimals is considered.

Figure \ref{fig:single} shows the results for the cases of $t_{\rm pl} $ = 2.0 Myr and $ t_{\rm pl}$ = 2.5 Myr on the left and the right panels, respectively.
We first discuss the case of $t_{\rm pl} $ = 2.0 Myr.
The top-left panel depicts temperature evolutions at the positions of $r$ = 0 km, 80 km, and 90 km.
Initially, the temperature increases with time, as a result of the continuous energy input from radionuclides $^{26}$Al.
The central region in the planetesimal ($r$ = 0 km) achieves the highest temperature, which exceeds 1000$^\circ$C at $t$ = 5.5 Myr.
On the other hand, the temperature at $r$ = 80 km ($r$ = 90 km) reaches 920 $^\circ$C (680 $^\circ$C) at $t$ = 5.5 Myr (4.2 Myr).
And then it starts to decrease because the heating rate due to decay of $^{26}$Al becomes lower than the cooling rate at the position.
This means that each position of the planetesimal has a peak temperature ($T_{\rm peak}(r)$) at $t = t_{\rm peak}(r)$.
The bottom-left panel of Figure \ref{fig:single} shows the temperature evolution at all the positions of the planetesimal.
We can see that the temperatures at $r \leq$ 65 km follow the same evolution and can increase to more than 1000 $^\circ$C.

For the case of $t_{\rm pl} $ = 2.5 Myr (see right panels of Figure \ref{fig:single}),
$T_{\rm peak}(r)$ at the center is 600 $^\circ$C, demonstrating that the planetesimal formed later has the lower $T_{\rm peak}$ at the same position than the earlier formed one.
This indicates that the thermal evolution is significantly affected by the formation time of planetesimals.

\subsection{The maximum temperature and the corresponding mass fraction} \label{res:single2}

\begin{figure*}
\plottwo{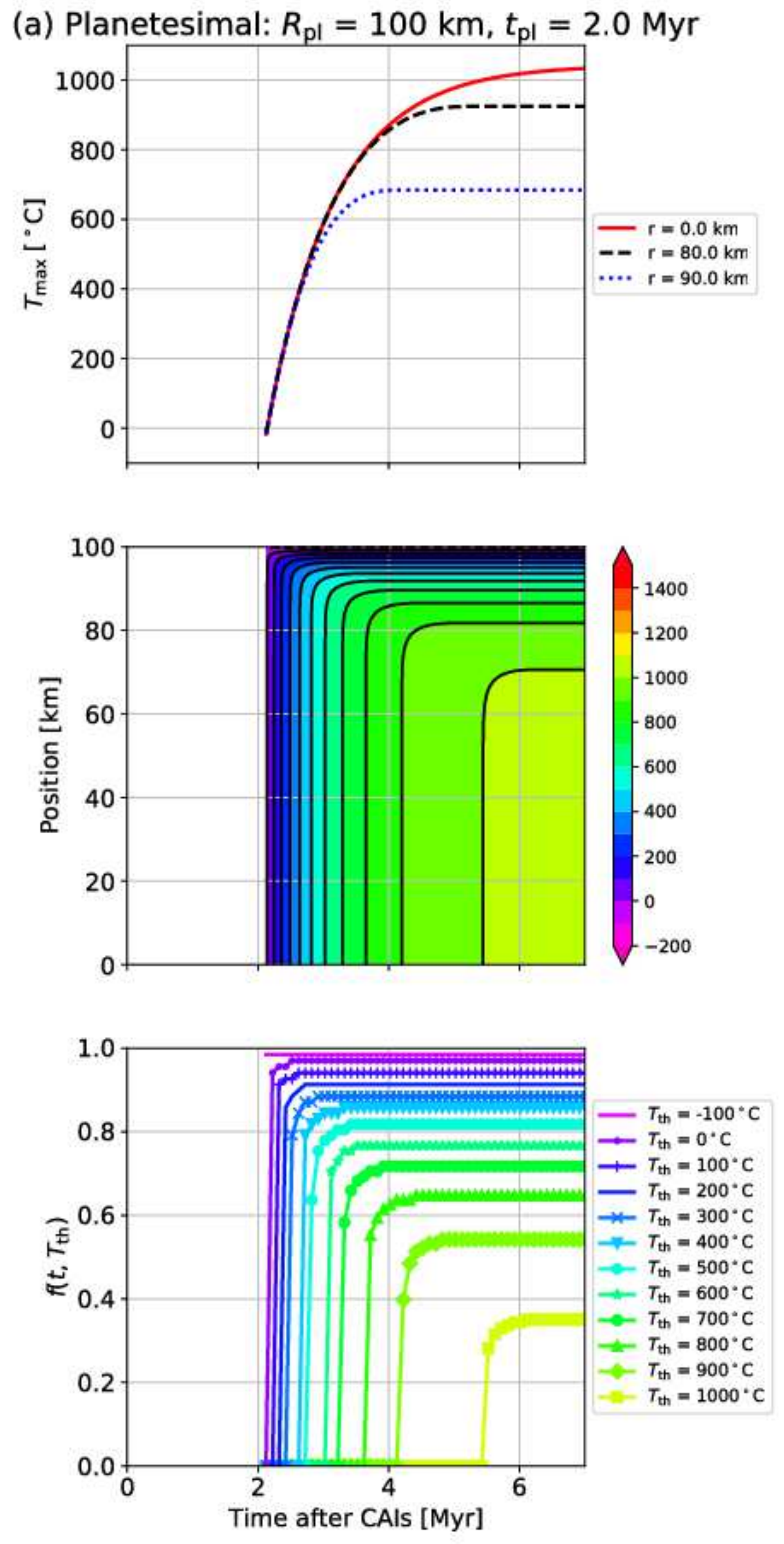}{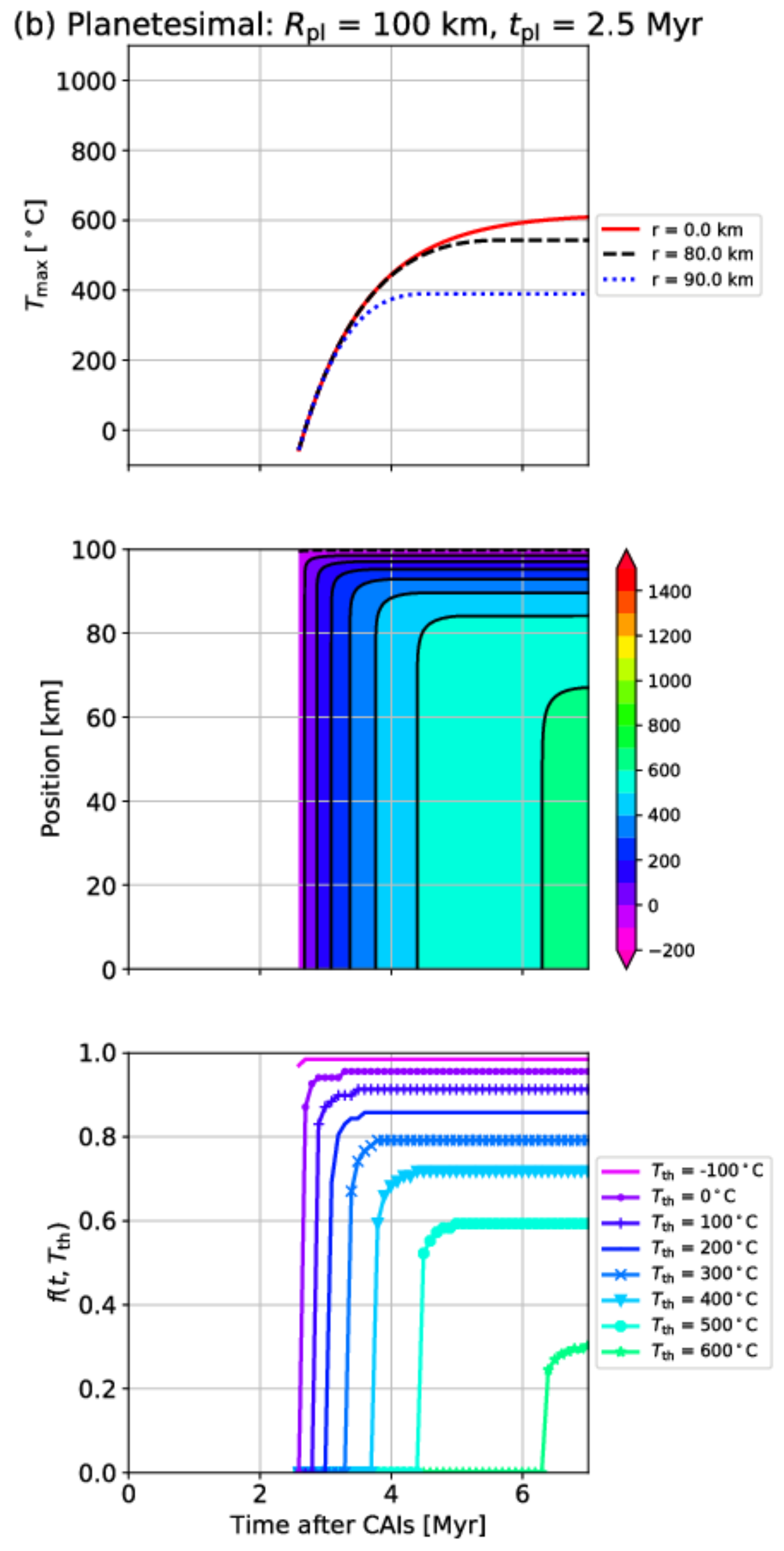}
\caption{Maximum temperatures of single-sized planetesimals ($R_{\rm pl}$ = 100 km) formed at (a) $t_{\rm pl}$ = 2.0 Myr and (b) $t_{\rm pl}$ = 2.5 Myr. 
Top panels show time evolutions of $T_{\rm max}(r,t)$ at positions of $r$ = 0 km (red solid lines), $r$ = 80 km (black dashed lines), and $r$ = 90 km (blue dotted lines), respectively.
Middle panels show $T_{\rm max}(r,t)$ evolutions at each position of planetesimals.
Bottom panels show the mass fraction of the planetesimals, $ f(t, T_{\rm th})$, that experienced the maximum temperatures higher than $T_{\rm th}$, defined in Equation (\ref{eq:smallf}). 
\label{fig:single2}}
\end{figure*}

As discussed above, the peak temperatures inside planetesimals that are heated by radioactive nuclei are different at different locations.
For the analysis that will be developed for understanding the origin of petrologic type of ordinary chondrites, we manipulate the computed values of the temperature ($T(r,t)$) as follows.

In this work, we are interested in the maximum temperatures that planetesimals experience through thermal evolution.
Given that the peak temperatures ($T_{\rm peak} (r)$) at position $r$ are achieved at different time ($t = t_{\rm peak} (r)$), 
the maximum temperature  ($T_{\rm max} (r,t)$) is described as
\begin{equation}
T_{\rm max} (r,t) =
                                    \left\{ \begin{tabular}{@{}l@{}} 
                                                                       $T(r,t)$                   at $t < t_{\rm peak}(r)$  \\
                                                                       $T_{\rm peak} (r)$ at $t \ge t_{\rm peak}(r)$, \\
                                             \end{tabular} \right.
                                             \label{eq:tmax}
\end{equation}
from the temperature evolution as given in Figure \ref{fig:single}.
The time evolutions of $T_{\rm max}(r,t)$ are shown in the top and middle panels of Figure \ref{fig:single2}.
As is obvious from Equation (\ref{eq:tmax}), $T_{\rm max}$ becomes constant after $t=t_{\rm peak}=$ 5.5 Myr (4.2 Myr) at $r$ = 80 km ($r$ =90 km) for $t_{\rm pl} = 2.0$ Myr.

We then introduce the mass fraction ($f(t, T_{\rm th})$) of planetesimals that is determined by a threshold temperature ($T_{\rm th}$).
When the single generation of planetesimals with the single size is considered,
a fraction of mass that experiences the temperatures equal to and higher than $T_{\rm th}$ can be computed as
\begin{equation}
 f(t, T_{\rm th}) = \frac{N_1}{M_1} \frac{4 \pi \rho}{3} r^3(T_{\rm max} (t) \geq T_{\rm th}), \label{eq:smallf}
\end{equation} 
where $t$ is explicitly labeled both in $f$ and $T_{\rm max}$ for clearly representing that both quantities are functions of time.

The bottom panels of Figure \ref{fig:single2} show the time evolutions of $f(t, T_{\rm th})$ for different $T_{\rm th}$.
Our results show that $f(t, T_{\rm th})$ with higher $T_{\rm th}$ rises up as time goes later.
This is the direct reflection that the temperature at each position increases with time (see the middle panels of Figure \ref{fig:single2}).
Then, any values of $f(t, T_{\rm th})$ become almost constant after $t \simeq$ 6 Myr for $t_{\rm pl}$ = 2.0 Myr.
This is because, at this time, almost all positions already pass through $T_{\rm peak}$, and $T_{\rm max}$ no longer changes.
Also, our results clearly show that a mass fraction of the planetesimals which undergo a high temperature is reduced for ones formed at later times 
(see the bottom-right panel of Figure \ref{fig:single2}):
$f(t, T_{\rm th} =600 ^\circ$C) = 0.3 and $f(t, T_{\rm th} = 800^\circ$C) = 0 at $t \ga 7$ Myr for the planetesimal formed at $t_{\rm pl}$ = 2.5 Myr.
On the contrary, the planetesimal formed at $t_{\rm pl}$ = 2.0 Myr has $f(t, T_{\rm th} =600 ^\circ$C) = 0.76 and $f(t,T_{\rm th} = 800^\circ$C) = 0.65 at $t \ga 7$ Myr.

This clearly demonstrates that a mass fraction of planetesimals that undergo the temperature above 600 $^\circ$C strongly depends on the formation time of planetesimals.

\subsection{Multiple generations of planetesimals with the single size} \label{res:muti1}

\begin{figure}
\plotone{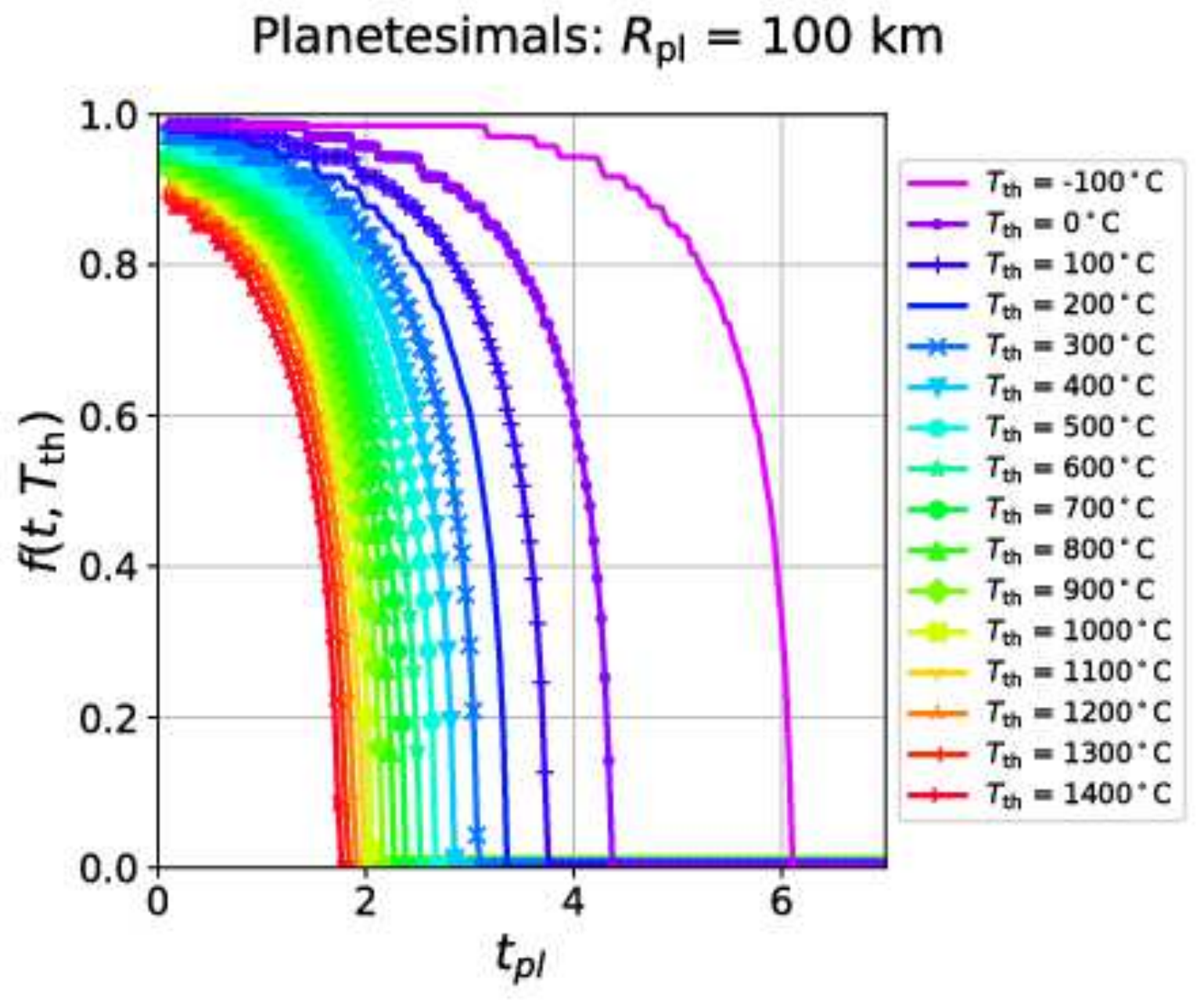}
\caption{Mass fraction $f(t, T_{\rm th})$, defined in Equation (\ref{eq:smallf}), that experienced $T_{\rm max}(t) \geq T_{\rm th}$ and is measured at $t_{\rm ms}$ = 7.0 Myr
for planetesimals which have radii of $R_{\rm pl}$ = 100 km and form at $t_{\rm pl}$ = 0.1 Myr to 7.0 Myr. \label{fig:multi}}
\end{figure}

We have so far considered a single generation of planetesimals, that is, $t_{\rm pl}= t_{\rm int} = t_{\rm fin}$.
In general, however, planetesimals are expected to form over some period. 
Here, we present the results from multiple generations of planetesimals ($t_{\rm int} \leq t_{\rm pl} \leq t_{\rm fin}$) with single sizes.
As an example, we adopt the following set of model parameters: 
$R_{\rm pl}$ = 100 km, $t_{\rm int} = 0.1$ Myr and $ t_{\rm fin} = 7.0$ Myr with a constant formation rate.

We first see the effect of planetesimal formation time on $f(t,T_{\rm th})$.
Figure \ref{fig:multi} plots $f(t, T_{\rm th})$ at $t$ = 7.0 Myr $\equiv t_{\rm ms}$ as a function of $t_{\rm pl}(t_{\rm int} \leq t \leq t_{\rm fin})$.
We confirm that the mass fraction becomes constant at $t_{\rm ms} \geq t_{\rm fin}$.
We find that as the formation time shifts later, $f(t_{\rm ms}, T_{\rm th})$ becomes lower. 
For instance, the planetesimals formed after $t_{\rm pl}$ = 2.3 Myr cannot reach 800 $^\circ$C.
Also, $f(t= t_{\rm ms},T_{\rm th}=0 ^\circ$C) = 0 at $t_{\rm pl} >$ 4.3 Myr, indicating that the planetesimals formed later than $t_{\rm pl}$ = 4.3 Myr cannot be heated above 0$^\circ$C even at the center.

\begin{figure}
\plotone{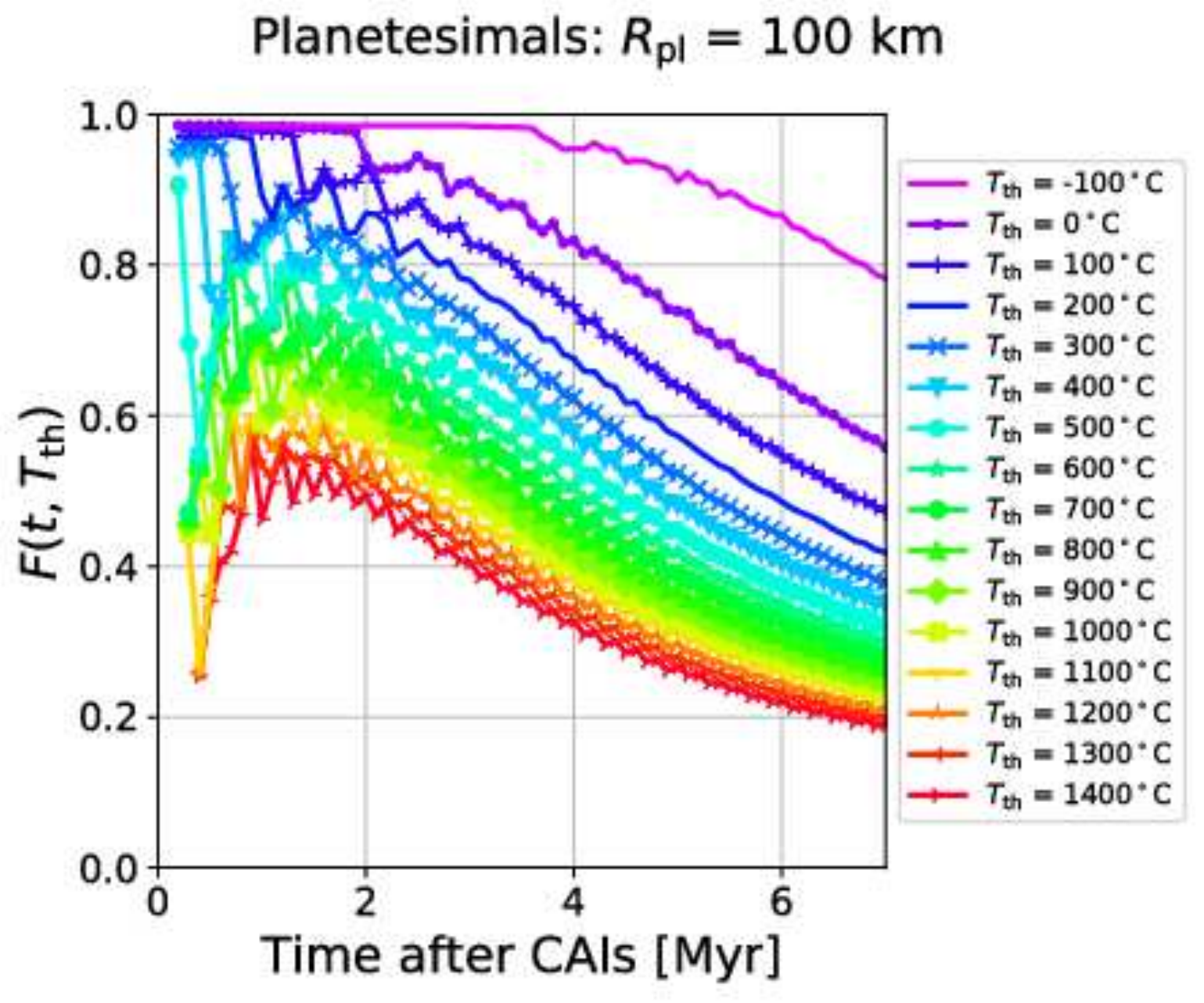}
\caption{Time evolution of mass fraction $F(t, T_{\rm th})$ defined in Equation (\ref{eq-fract}) 
for planetesimals with $R_{\rm pl}$ = 100 km which formed at $t_{\rm pl}$ = 0.1 Myr to 7.0 Myr. \label{fig-mass}}
\end{figure}

We then examine the behavior of time-integrated mass fractions of planetesimals formed at multiple epochs.
The fraction of mass $F(t, T_{\rm th})$ that experienced $T_{\rm max} \geq T_{\rm th}$ for all of the formed planetesimals is given as
\begin{equation}
 F(t, T_{\rm th})  = \frac{1}{M_2(t)}  \sum^t_{t^\prime = t_{\rm int}} \phi(t^\prime) \frac{4 \pi \rho}{3}  r^3(T_{\rm max} (t^\prime) \geq T_{\rm th}). \label{eq-fract} 
\end{equation} 

Figure \ref{fig-mass} shows the time evolution of $F(t, T_{\rm th})$.
It can be seen that $F(t, T_{\rm th})$ for $T_{\rm th} \ge 800 ^\circ$C increases until $t \simeq$ 1.5 Myr.
This stems from the fact that it takes about 1 Myr for earlier formed planetesimals to reach $T_{\rm peak}(r)$.
On the other hand, $F(t, T_{\rm th})$ for $T_{\rm th} \ge 800 ^\circ$C starts to decreases after $t$ = 1.5 Myr because a small or no part of later formed planetesimals can reach higher temperatures than 800 $^\circ$C (also see Figure \ref{fig:multi}). 
Our results also show some spiky structures at the early times.
These are numerical noises, but do not affect our conclusions that are derived from the later stage of the evolutions.

\begin{figure}
\plotone{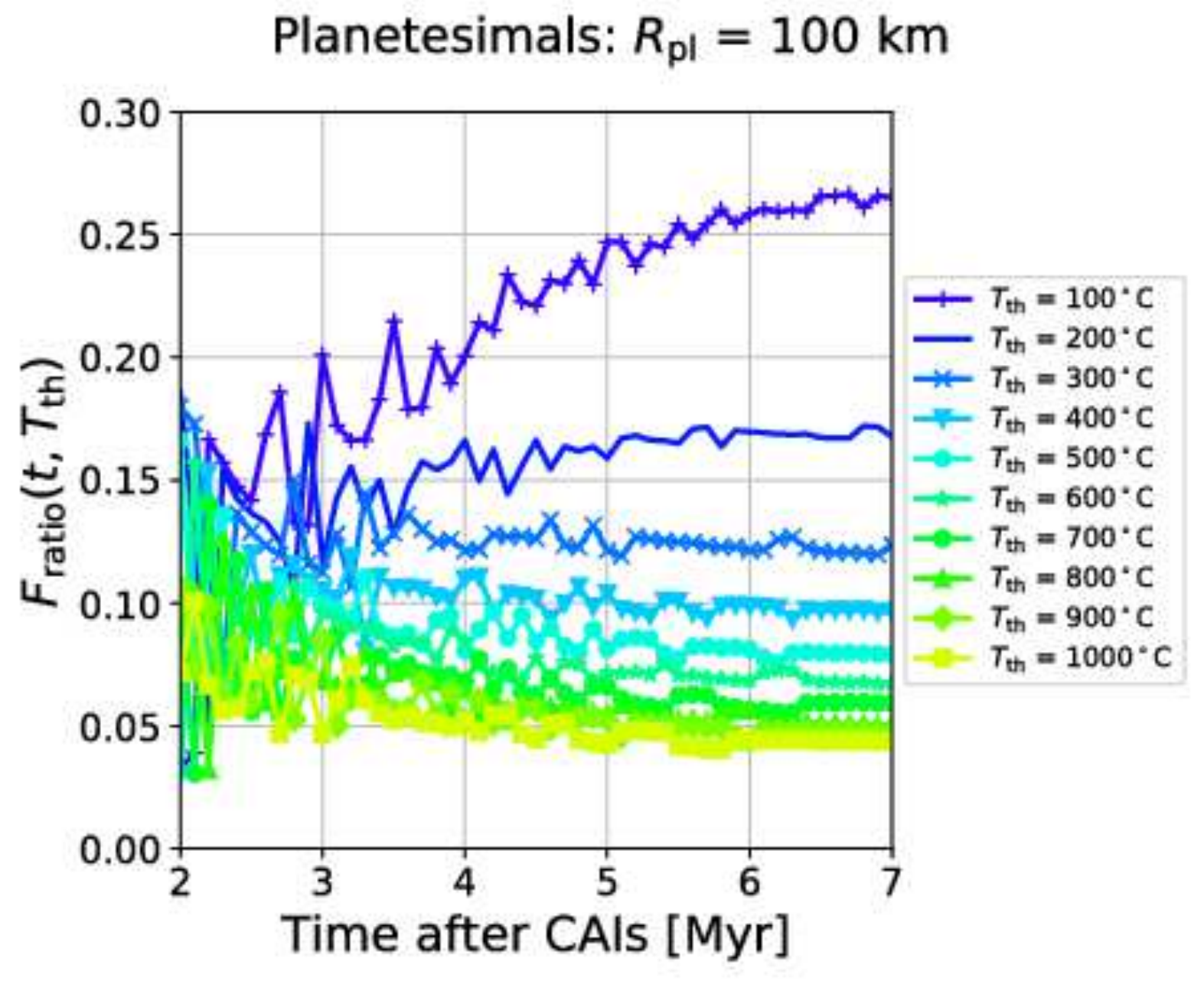}
\caption{Time evolution of fraction ratio $F_{\rm ratio} (t, T_{\rm th})$ defined in Equation (\ref{eq:ratio}) 
for planetesimals of $R_{\rm pl}$ = 100 km which formed at $t_{\rm pl}$ = 0.1 Myr to 7.0 Myr. \label{fig-type-single}}
\end{figure}

In this study, we examine the abundance of ordinary chondrites moderately metamorphosed in thermally evolving planetesimals.
In order to perform direct comparison among our results, we define the ratio of $F(t, T_{\rm th})$ as follows,
\begin{equation}
F_{\rm ratio} (t, T_{\rm th})  \equiv \frac{F(t, T_{\rm th}-100 ^\circ{\rm C}) - F(t, T_{\rm th})}{F(t, T_{\rm th}=0 ^\circ {\rm C}) - F(t, T_{\rm th}= 1000 ^\circ {\rm C})} \label{eq:ratio},
\end{equation}
where $T_{\rm th}$ varies from 100 $^\circ$C to 1000 $^\circ$C.

Figure \ref{fig-type-single} shows the time evolutions of $F_{\rm ratio}(t, T_{\rm th})$ for different $T_{\rm th}$.
We find that $F_{\rm ratio} (t, T_{\rm th})$ becomes almost constant after $t$ = 5.0 Myr.
Planetesimals formed later than 5.0 Myr cannot contribute to the change of $F_{\rm ratio}(t, T_{\rm th})$, because they do not experience the temperature higher than 100 $^\circ$C.
Accordingly, we focus on planetesimals that form earlier than 5.0 Myr in the following sections,
and denote $F_{\rm ratio} (t = t_{\rm ms}$ = 5.0 Myr, $T_{\rm th})$ as $F_{\rm ratio} (T_{\rm th})$ for brevity.

\begin{figure}
\plotone{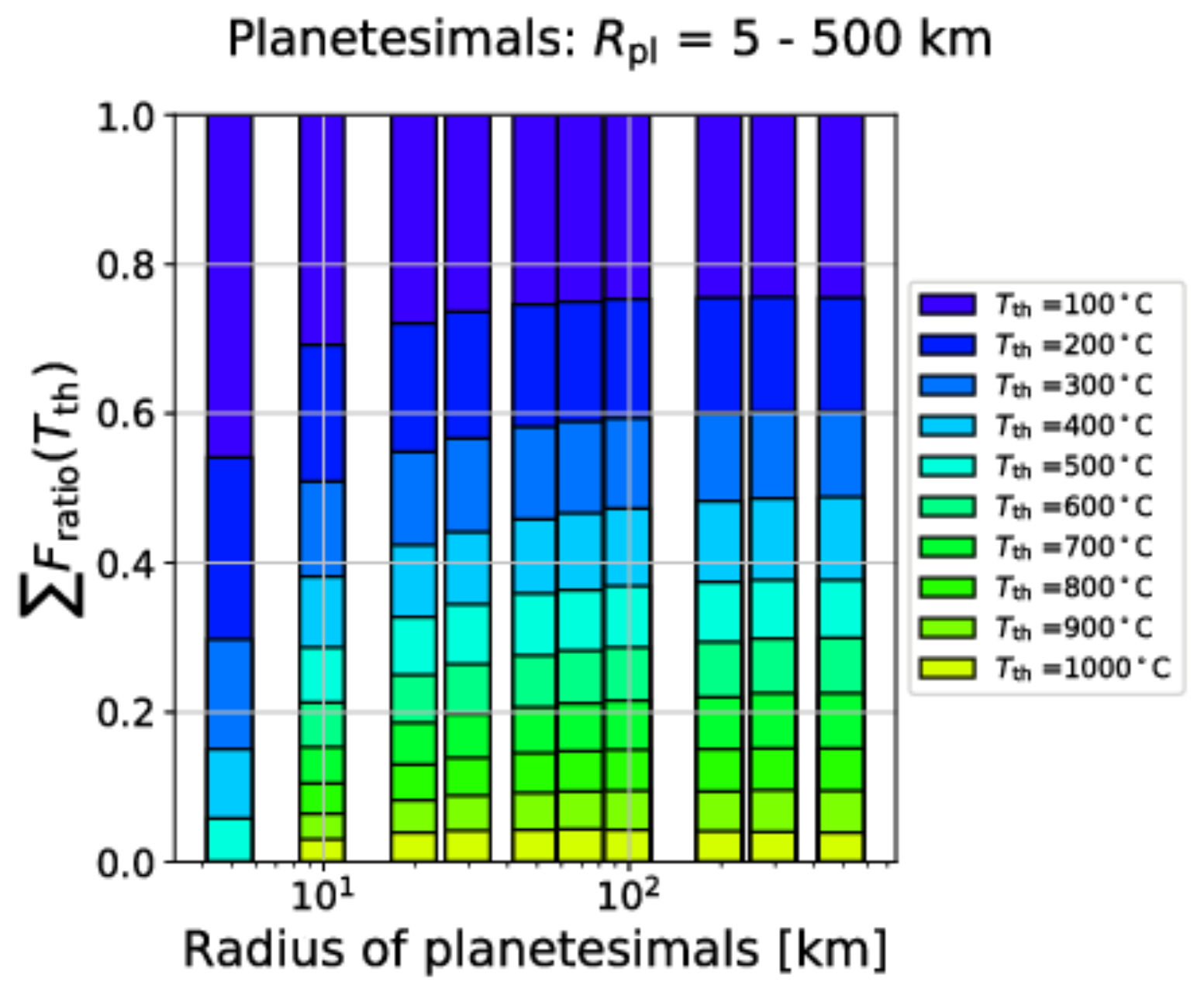}
\caption{Cumulative representation of $F_{\rm ratio} (T_{\rm th})$ defined in Equation (\ref{eq:ratio}) for planetesimals with the radius of 5, 10, 20, 30, 50, 70, 100, 200, 300, and 500 km. 
The planetesimals form from 0.1 Myr to 5.0 Myr with a constant formation rate.
\label{fig-radius-bar}}
\end{figure}

We are now in a position to examine the effect of $R_{\rm pl}$ on $F_{\rm ratio} (T_{\rm th})$, which is shown in Figure \ref{fig-radius-bar}.
In this figure, we adopt the following parameters:
$R_{\rm pl}$ = 5 km - 500 km, $t_{\rm int} = 0.1$ Myr, $t_{\rm fin} = 5.0$ Myr, and the constant formation rate is considered.
We find that $F_{\rm ratio} (T_{\rm th})$ for larger planetesimals with $R_{\rm pl} \geq $ 100 km are very similar,
and $F_{\rm ratio} (T_{\rm th} \geq 600^\circ {\rm C})$ is 0.29.
This means that even if we consider the arbitrary size distribution for planetesimals larger than 100 km, the results do not change very much.
On the other hand, $F_{\rm ratio} (T_{\rm th})$ of planetesimals with $R_{\rm pl} \leq $ 50 km are different.
Planetesimals of $R_{\rm pl} = $ 50 km have $F_{\rm ratio} (T_{\rm th} \geq 600 ^\circ {\rm C})$ of 0.27.
For $R_{\rm pl} =$ 10 km and 20 km, $F_{\rm ratio} (T_{\rm th} \geq 600^\circ {\rm C)}$ are 0.21 and 0.25, respectively.
This indicates that smaller planetesimals produce lower $F_{\rm ratio} (T_{\rm th} \geq 600^\circ {\rm C})$ than larger ones, reflecting the fact that they cannot reach higher temperatures than the larger ones due to more effective cooling from their surfaces.

It should be stressed that radius of planetesimal ($R_{\rm pl}$) plays an important role in thermal evolution of planetesimals and hence the resulting ratio ($F_{\rm ratio} (T_{\rm th})$).
Also, the results are not affected by planetesimals that form later than 5.0 Myr.

\subsection{Multiple generations of planetesimals with the size distribution} \label{res:size}

In this section, we present the results that are obtained from the most realistic case 
where planetesimals form within the time interval between $t=t_{\rm int}$ and $t=t_{\rm fin}$ and they have the size distribution from $R_{\rm pl, min}$ to $R_{\rm pl, max}$.

Under this situation, the mass fraction of planetesimals that experience a threshold temperature ($T_{\rm th}$) or higher is written as
\begin{eqnarray}
  \label{eq:largef}
 \mathscr{F}(t, T_{\rm th})  & =          &  \frac{1}{M_3(t)}  \sum^t_{t^\prime = t_{\rm int}} \phi (t^\prime)   \\
                                           & \times  & \int_{R_{\rm pl, min}}^{R_{\rm pl, max}} dR_{\rm pl} n_{\rm pl} (R_{\rm pl}) \frac{4 \pi \rho}{3}   r^3(T_{\rm max} (t^\prime) > T_{\rm th}, R_{\rm pl}).  \nonumber \label{eq:fratio}
\end{eqnarray} 
Note that $r$ is now a function of $R_{\rm pl}$.
Then, the ratio of $\mathscr{F}(t, T_{\rm th})$ can be given as
\begin{equation}
\label{eq:ratio2}
\mathscr{F}_{\rm ratio} (t, T_{\rm th})  \equiv \frac{\mathscr{F}(t, T_{\rm th}-100 ^\circ{\rm C}) - \mathscr{F}(t, T_{\rm th})}
                                                                              {\mathscr{F}(t, T_{\rm th}= 0 ^\circ {\rm C}) - \mathscr{F}(t, T_{\rm th}= 1000 ^\circ {\rm C})},                                 
\end{equation}
for $100 ^\circ {\rm C} \leq T_{\rm th} \leq 1000 ^\circ {\rm C}$.
Armed with this formulation, we carry out a parameter study and examine how $\mathscr{F}_{\rm ratio} (t, T_{\rm th})$ at $t=t_{\rm ms}$ is influenced by model parameters of planetesimals.
For brevity, $\mathscr{F}_{\rm ratio} (t = t_{\rm ms}$ = 5.0 Myr, $T_{\rm th})$ is denoted as $\mathscr{F}_{\rm ratio} (T_{\rm th})$.
Unless otherwise mentioned, we take the power law index $\alpha$ = 2.8 and consider that planetesimals form from $t_{\rm int} = 0.1$ Myr up to $t_{\rm fin}$ = 5.0 Myr with a constant formation rate.

\subsubsection{Dependence on size ranges of planetesimals} \label{res:size2}

\begin{figure}
\plotone{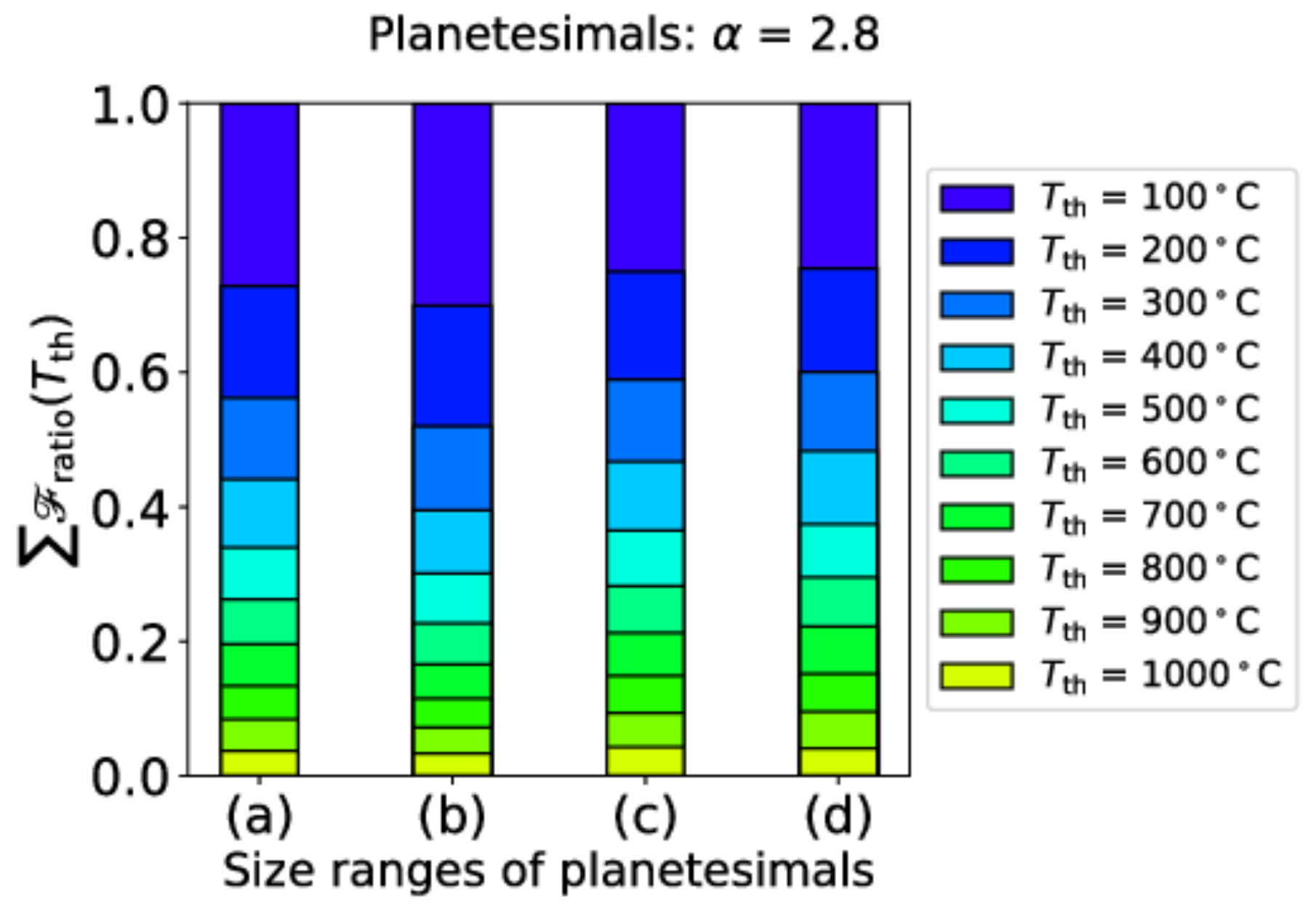}
\caption{Cumulative representation of $\mathscr{F}_{\rm ratio} (T_{\rm th})$ defined in Equation (\ref{eq:ratio2}) 
for planetesimals with size ranges of (a) all (1 km - 500 km), (b) small (1 km - 50 km), (c) middle (60 km - 90 km), and (d) large (100 km - 500 km). 
The power law index is $\alpha$ = 2.8 and planetesimals form from 0.1 Myr to 5.0 Myr with a constant formation rate.
\label{fig-range-bar}}
\end{figure}

Planetesimals are likely to form with some size range (size distribution) but not with a single size. 
Therefore, we consider four size ranges; small (1 km - 50 km in radius), middle (60 km - 100 km in radius), 
large (100 km - 500 km in radius), and all (1 km - 500 km in radius).

Figure \ref{fig-range-bar} shows the resulting values of  $\mathscr{F}_{\rm ratio} (T_{\rm th})$ for the four size ranges.
We find that the results of the middle and large size ranges ((c) and (d) in Figure \ref{fig-range-bar}) look similar. 
Both of them have $\mathscr{F}_{\rm ratio} (T_{\rm th} \geq 600 ^\circ {\rm C})$ = 0.29.
On the other hand, the small size range (Figure \ref{fig-range-bar} (b)) takes a smaller value of $\mathscr{F}_{\rm ratio} (T_{\rm th} \geq 600 ^\circ{\rm C})$ = 0.22.
As we see in the previous section, only small portion of the planetesimal can reach higher temperatures in small sized planetesimals.
Figure \ref{fig-range-bar}(a) shows the ratio of all planetesimals combined from other three ranges.

From this, we can conclude that the cumulative mass fraction ratio of planetesimals that experience certain temperatures depends on their size range. 
In particular, there is a significant difference between the smaller size range ($R_{\rm pl} \leq $ 50 km) and larger one ($R_{\rm pl} \geq$ 60 km).

\subsubsection{Dependence on onset of planetesimal formation} \label{res:time}

\begin{figure*}
\gridline{\fig{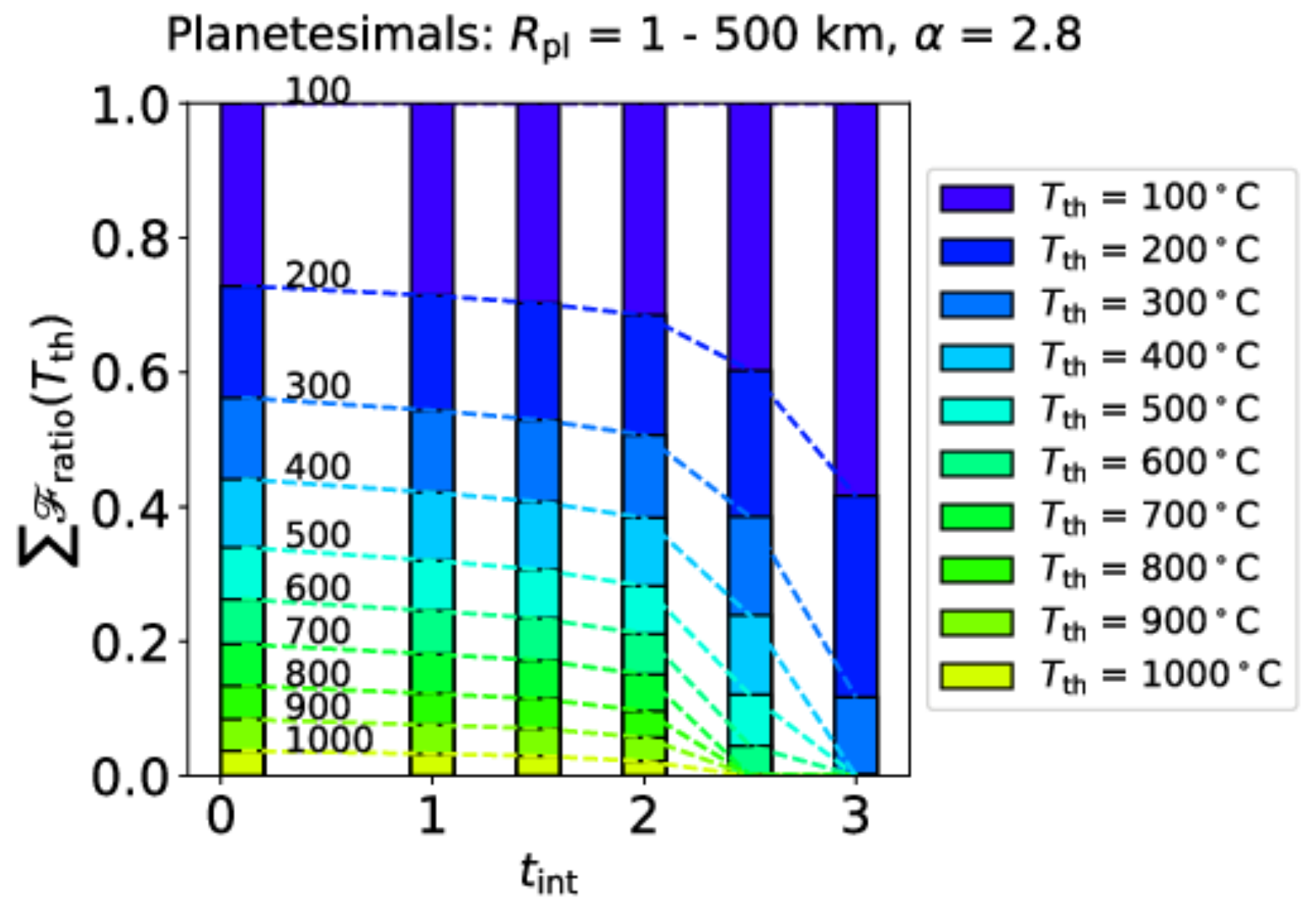}{0.5\textwidth}{(a)}
\fig{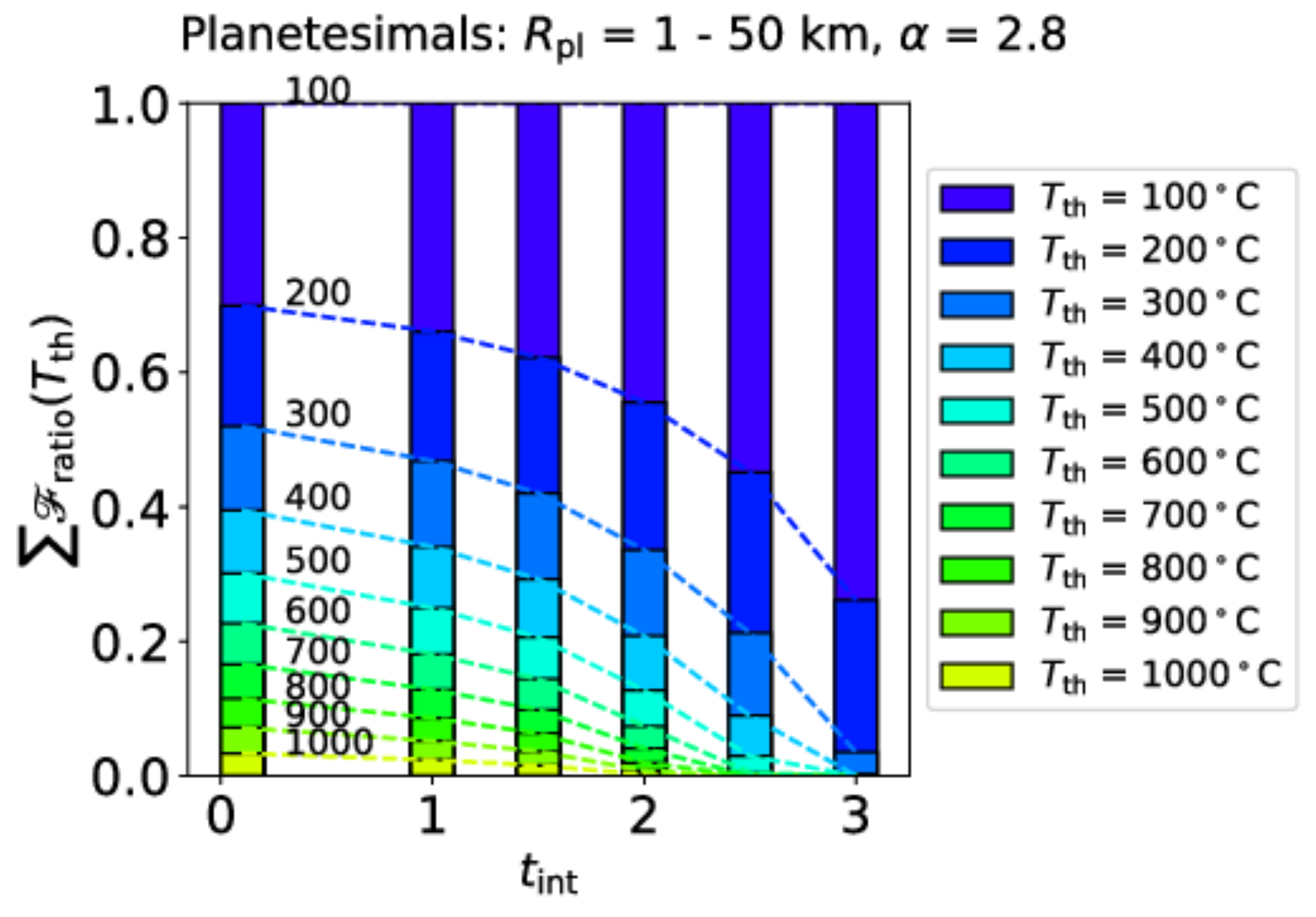}{0.5\textwidth}{(b)}}
\gridline{\fig{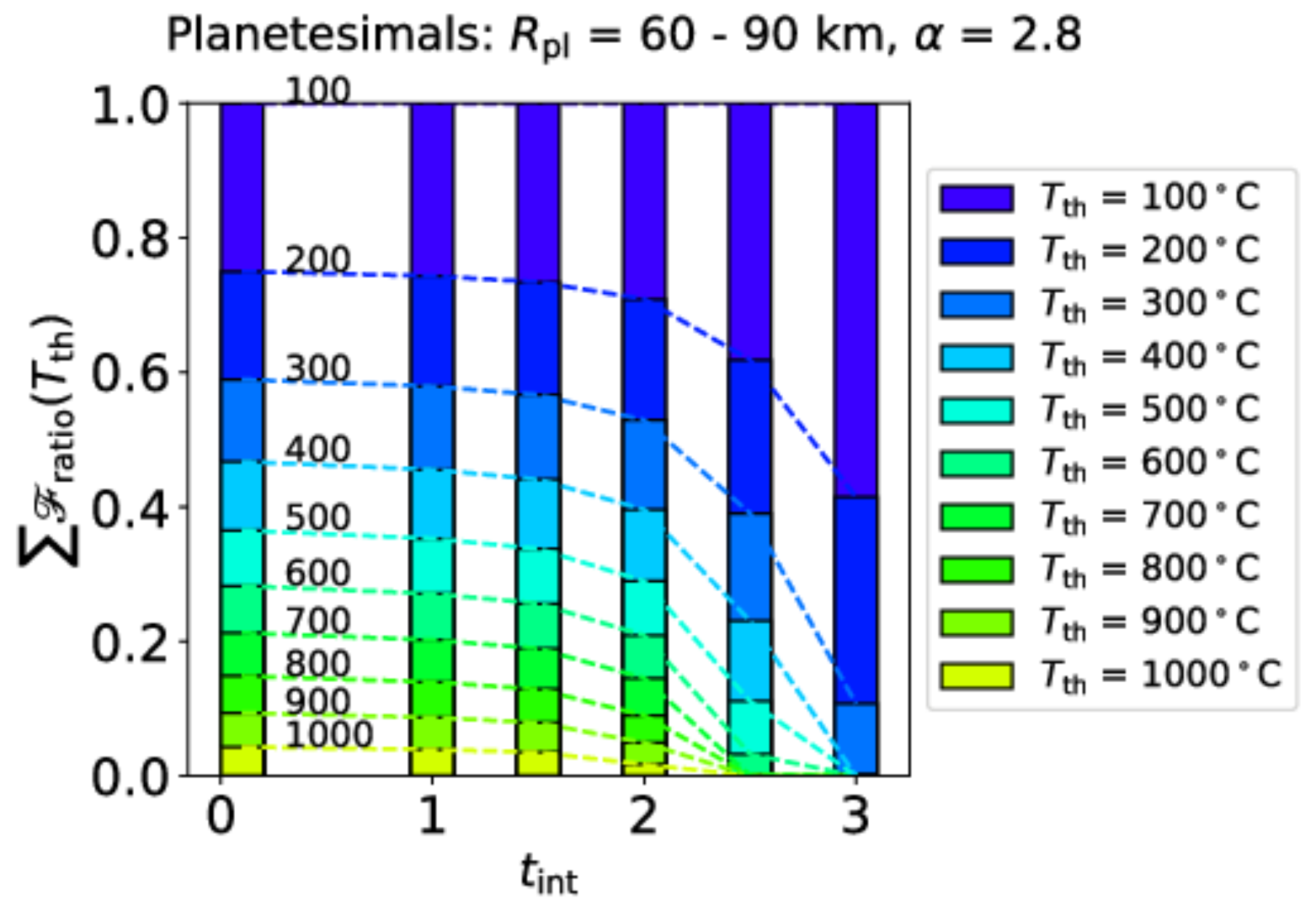}{0.5\textwidth}{(c)}
\fig{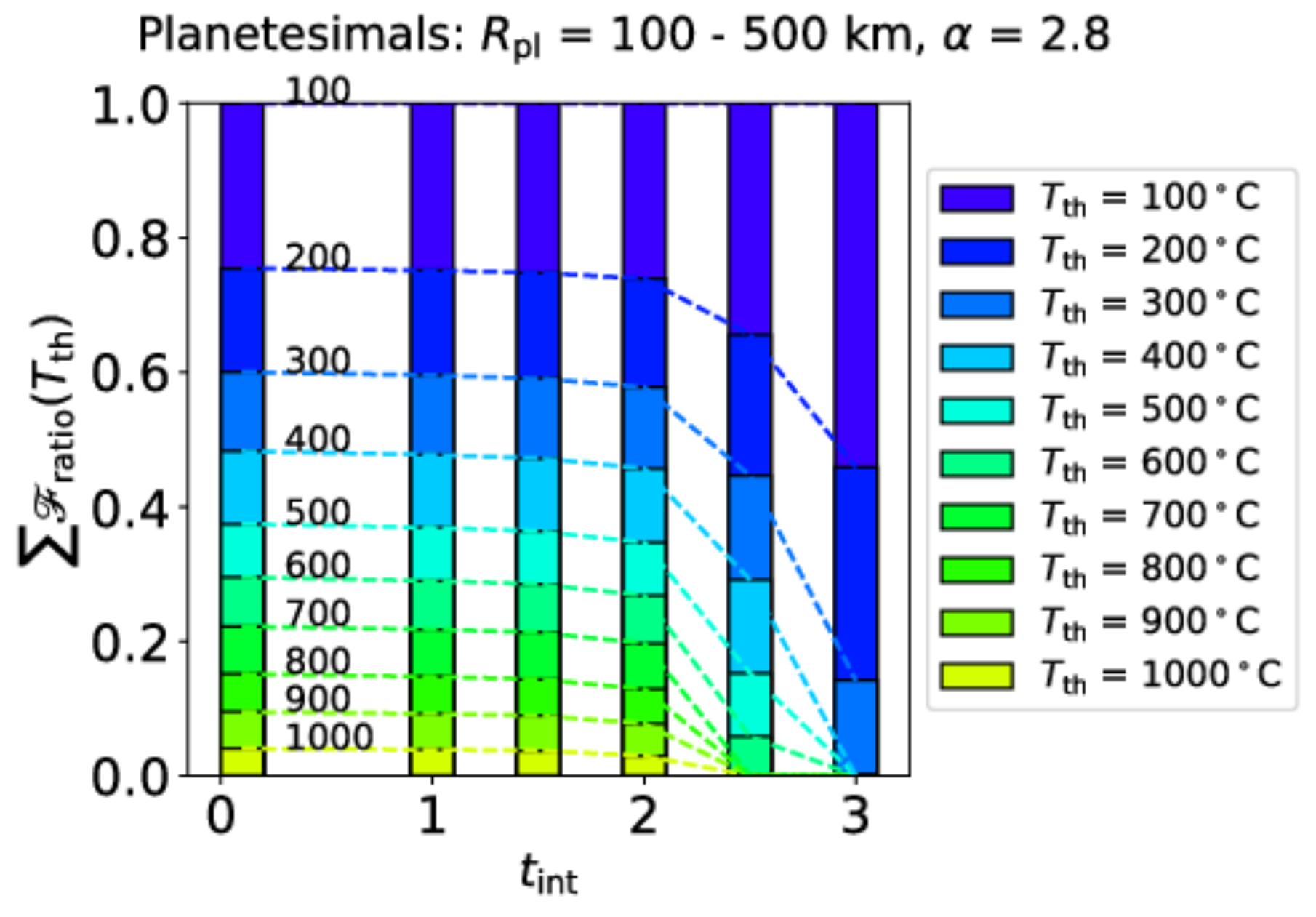}{0.5\textwidth}{(d)}}
\caption{Cumulative representation of $\mathscr{F}_{\rm ratio} (T_{\rm th})$ defined in Equation (\ref{eq:ratio2}) 
for planetesimals with size ranges of (a) all (1 km - 500 km), (b) small (1 km - 50 km), (c) middle (60 km - 90 km), and (d) large (100 km - 500 km). 
The results are shown for the onset of planetesimal formation $t_{\rm int}$ taken at 0.1, 1.0, 1.5, 2.0, 2.5, and 3.0 Myr.
The power law index is $\alpha$ = 2.8 and planetesimals form with a constant formation rate.
\label{fig-size-range}}
\end{figure*}

One of our aims in this paper is to constrain the formation time of planetesimals.
In previous sections, we have considered that planetesimals start to form at $t_{\rm int} = 0.1$ Myr. 
In this subsection, we see the dependence of the results on the formation timing ($t_{\rm int}$) of planetesimals.

As shown in Figure \ref{fig:multi}, the mass fraction with higher temperatures decreases with increasing the formation rate.
If the formation time is later than 2.0 Myr after CAI formation, there is no chance to reach 1000 $^\circ$C.
Figure \ref{fig-size-range} shows $\mathscr{F}_{\rm ratio} (T_{\rm th})$ for the four size ranges with onset times of planetesimal formation,
$t_{\rm int}$ = 0.1, 1.0, 1.5, 2.0, 2.5, and 3.0 Myr.
When the onset time of planetesimal formation becomes later, $\mathscr{F}_{\rm ratio} (T_{\rm th} \geq 600 ^\circ {\rm C})$ decreases.
In particular, $\mathscr{F}_{\rm ratio} (T_{\rm th} \geq 600 ^\circ {\rm C})$ in small size range (1 km - 50 km) rapidly decreases as $t_{\rm int}$ goes later (Figure \ref{fig-size-range}(b)).
This is because smaller planetesimals cannot reach temperatures higher than 600 $^\circ$C when they form later
 ($t_{\rm pl}$ = 1.5 Myr for $R_{\rm pl}$ = 10 km and $t_{\rm pl}$ = 2.5 Myr for $R_{\rm pl}$ = 50 km).

We should emphasize that the onset time of planetesimal formation should be earlier than 2.2 Myr (1.2 Myr) to contain the volume 
which experiences $T_{\rm max}$ = 800 $^\circ$C 
for $R_{\rm pl} = 50$ km ($R_{\rm pl} = $ 10 km).
This is important to compare our results with fall statistics of ordinary chondrites (see \S\ref{sec:disc}).

\subsubsection{Dependence on size distributions of planetesimals} \label{res:dist}

\begin{figure}
\gridline{\fig{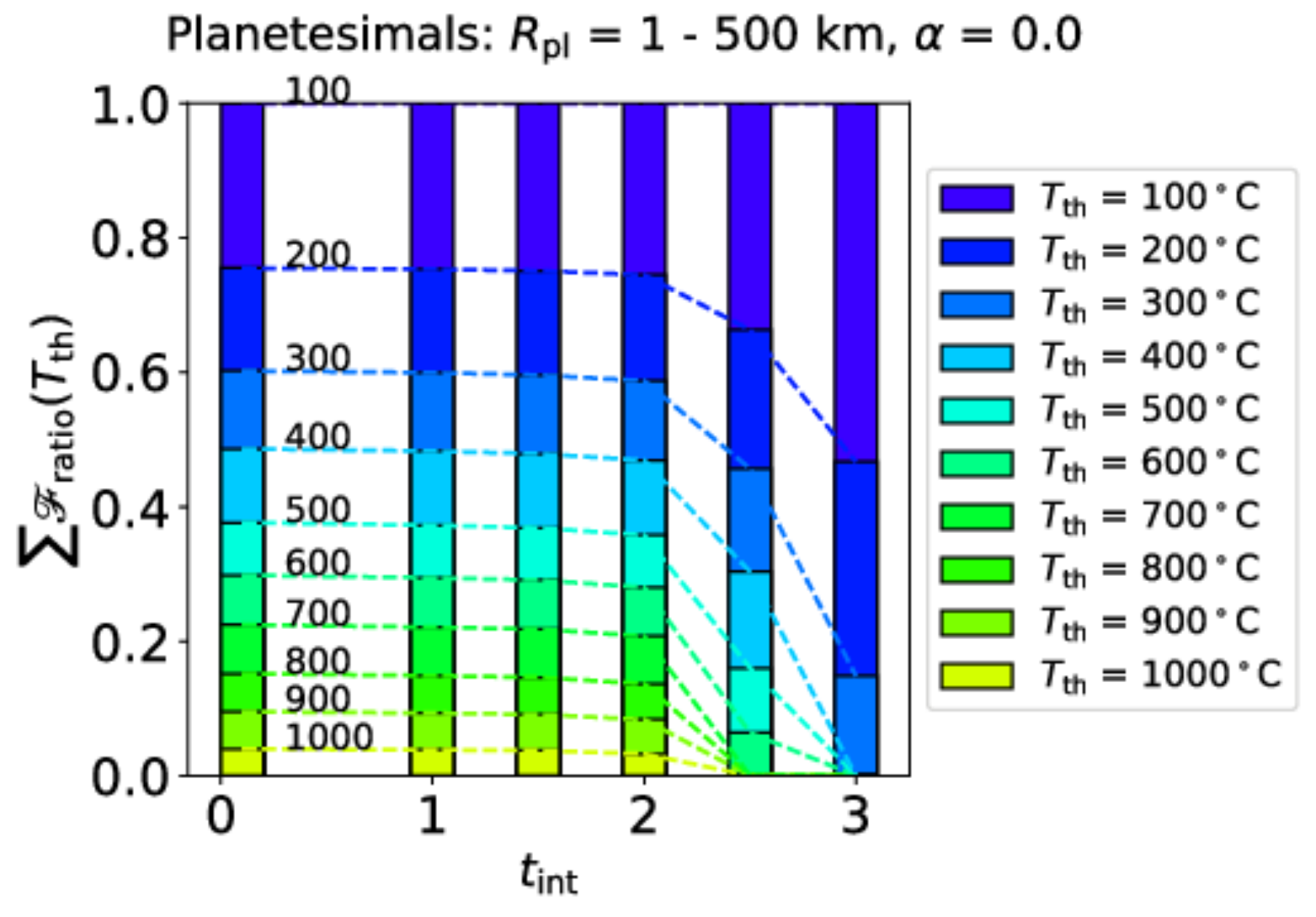}{0.5\textwidth}{(a)}}
\gridline{\fig{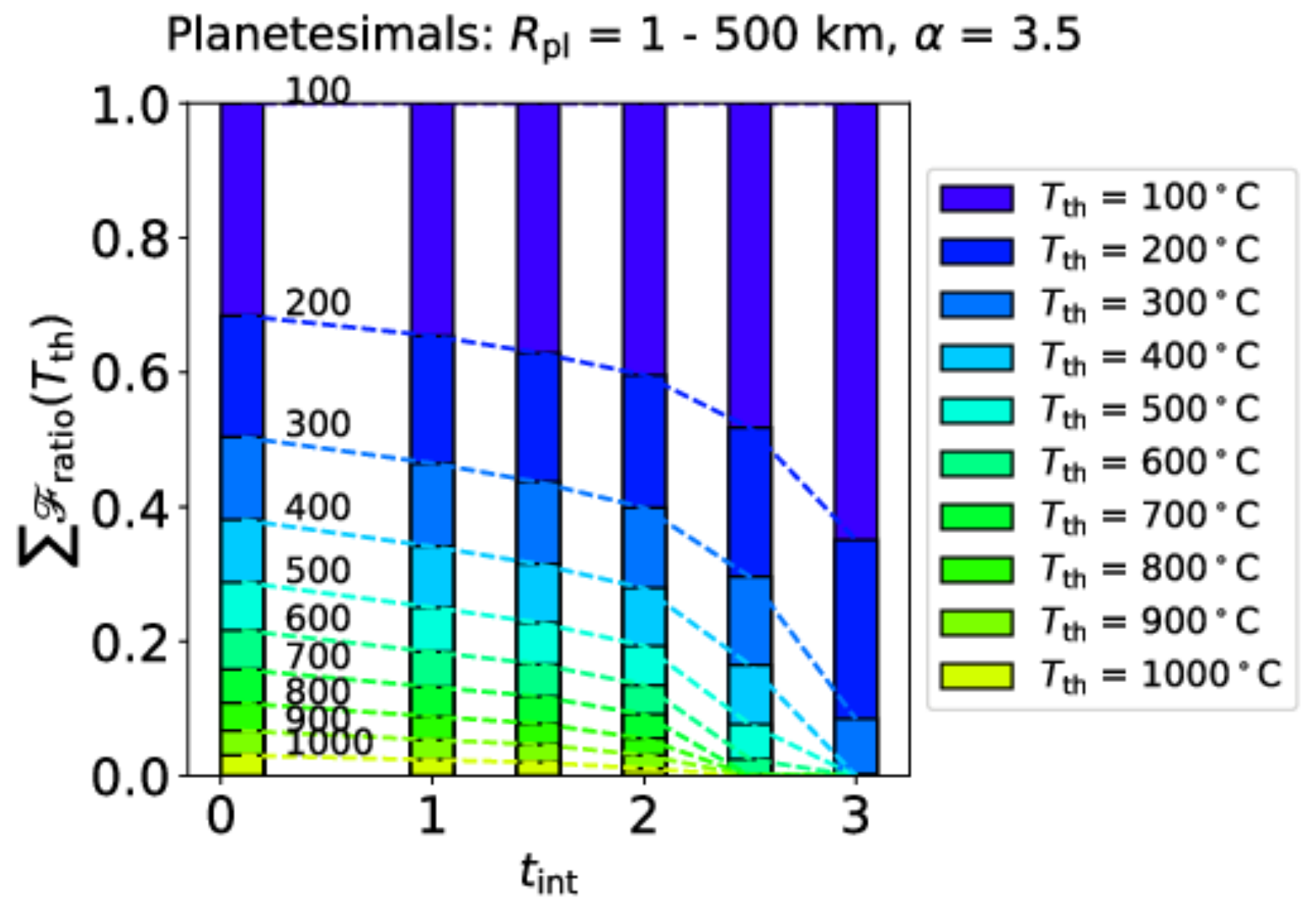}{0.5\textwidth}{(b)}}
\gridline{\fig{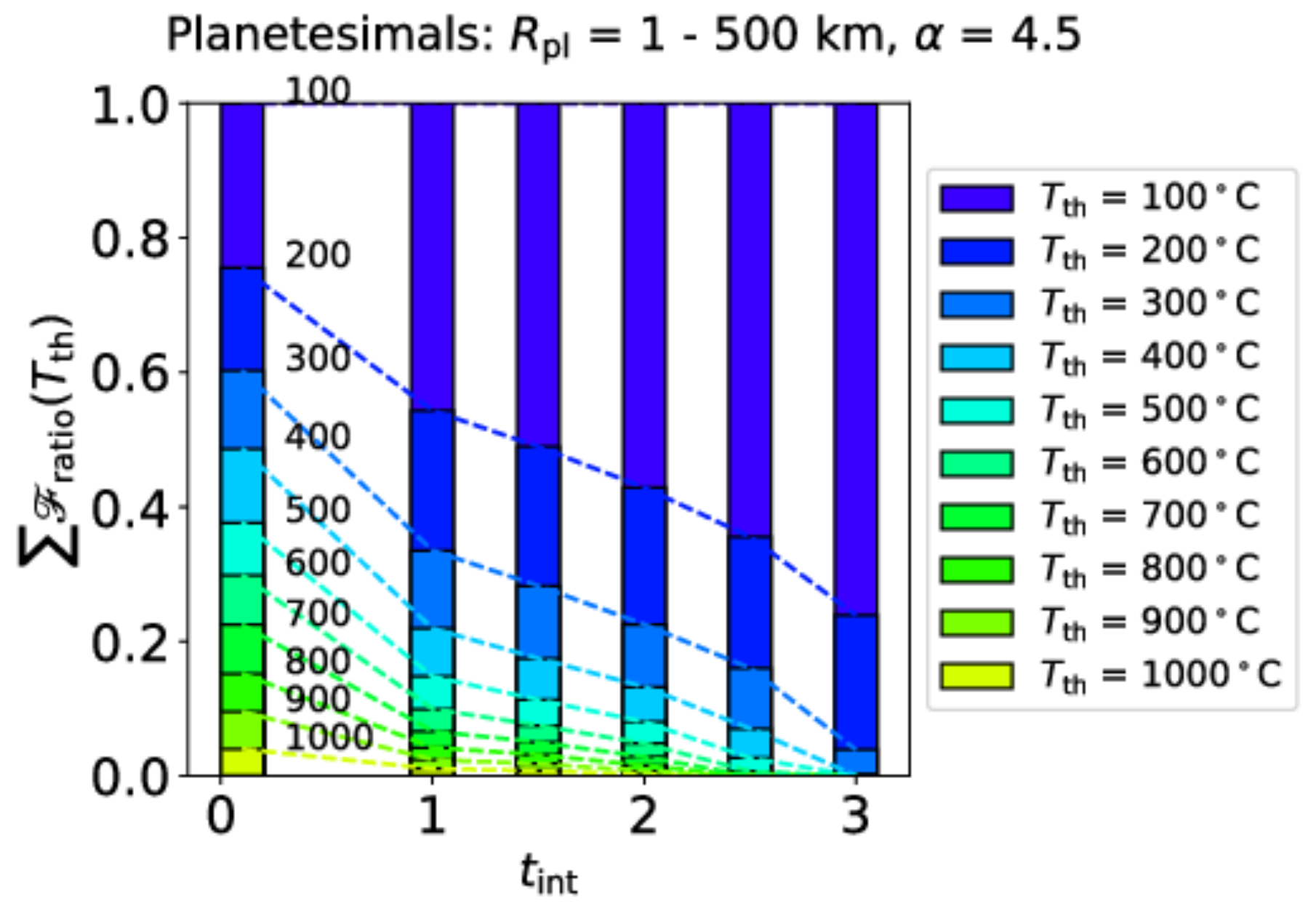}{0.5\textwidth}{(c)}}
\caption{Same as in Figure \ref{fig-size-range} (a) but different power law indexes (see \S\ref{res:dist}).
(a) $\alpha$ = 0.0, (b) $\alpha$ = 3.5, and $\alpha$ = 4.5. \label{fig-power-law}}
\end{figure}

Next we see the effect of the size distribution of planetesimals.
So far, we take $\alpha$ = 2.8 as the power-law index of size distribution.
This can be regarded as the initial size distribution of planetesimals \citep{jmb15,say16}.
Meanwhile, $\alpha$ =  3.5 is known to be the size distribution to explain the current asteroid belt \citep{mbn09}.
We also consider $\alpha$ = 0 and 4.5 as extreme cases.

The results for different size distributions with $\alpha$ = 0, 3.5, and 4.5 are shown in Figure \ref{fig-power-law}.
When we compare $\alpha$ = 2.8 (Figure \ref{fig-size-range} (a)) and 3.5 (Figure \ref{fig-power-law} (b)), we find a moderate difference between them:
$\mathscr{F}_{\rm ratio} (T_{\rm th} \geq 600 ^\circ{\rm C})$ at $t_{\rm int}$ = 2.0 Myr are 0.21 and 0.13 for $\alpha$ = 2.8 and 3.5, respectively. 
For $\alpha=0.0$, which means that the number of any size of planetesimals is the same,
$\mathscr{F}_{\rm ratio} (T_{\rm th})$ is mainly determined by larger planetesimals: that is $\mathscr{F}_{\rm ratio} (T_{\rm th} \geq 600 ^\circ{\rm C})$ = 0.28 at $t_{\rm int}$ = 2.0 Myr (Figure \ref{fig-power-law} (a)). 
This is comparable with $\mathscr{F}_{\rm ratio}(T_{\rm th} \geq 600 ^\circ{\rm C})$ = 0.27 at $t_{\rm int}$ = 2.0 Myr for $\alpha$ = 2.8 and $R_{\rm pl}$ = 100 km - 500 km
(Figure \ref{fig-size-range} (d)). 
When we take $\alpha$ = 4.5, smaller planetesimals regulate the result and $\mathscr{F}_{\rm ratio}(T_{\rm th} \geq 600 ^\circ{\rm C})$ is lower:
$\mathscr{F}_{\rm ratio} (T_{\rm th} \geq 600 ^\circ{\rm C})$ at $t_{\rm int}$ = 2.0 Myr is 0.05 for $\alpha$ = 4.5 at $t_{\rm int}$ = 2.0 Myr (Figure \ref{fig-power-law} (d)).

The value of $\mathscr{F}_{\rm ratio} (T_{\rm th})$ depends largely on the size distribution in the extreme case of $\alpha$ = 0.0 or 4.5.
However, the effect of the size distribution is moderate, as long as $\alpha \sim$ 3.0.

\subsubsection{Dependence on formation rate of planetesimals} \label{res:rate}

\begin{figure}
\gridline{\fig{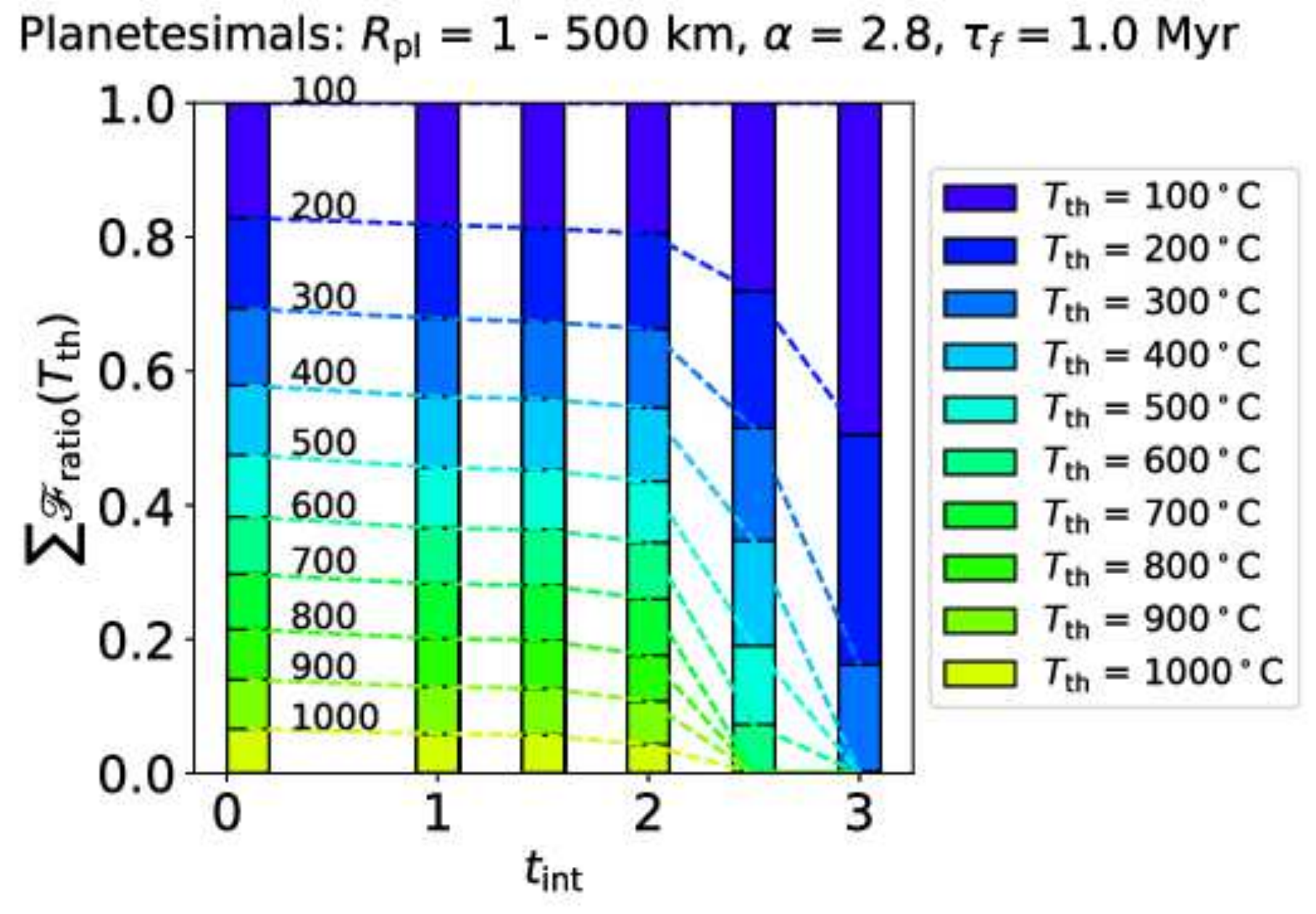}{0.5\textwidth}{(a)}}
\gridline{\fig{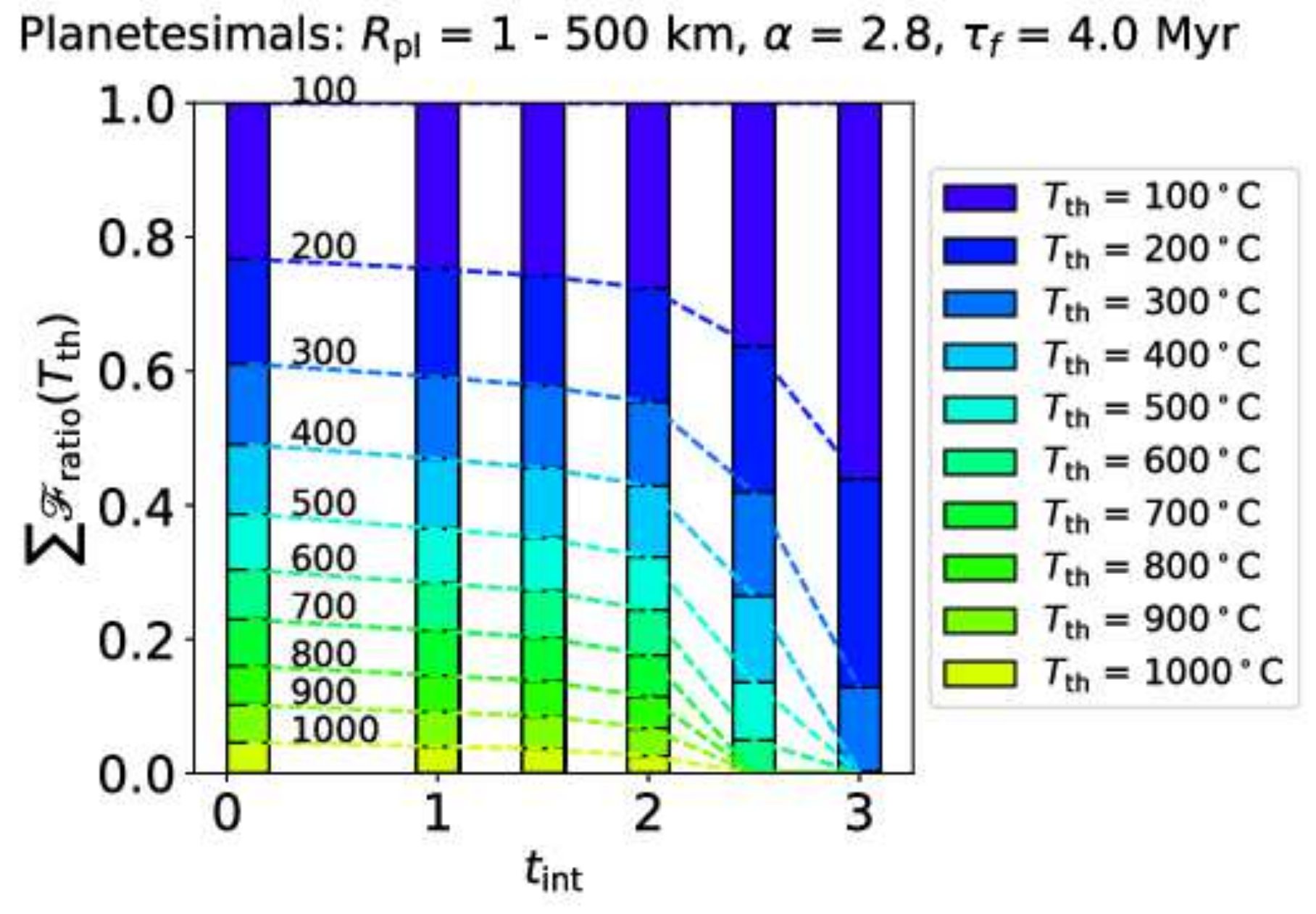}{0.5\textwidth}{(b)}}
\caption{Same as in Figure \ref{fig-size-range} (a) but different formation time scale $\tau_f$(see also \S\ref{res:rate}). 
(a) $\tau_f$ = 1.0 Myr and (b) $\tau_f$ = 4.0 Myr \label{fig-formation-rate}}
\end{figure}

In the previous sections, we have assumed that the formation rate of planetesimals is constant.
Here, we consider that the formation rate decreases on the timescale of $\tau_f$ (see Section \ref{sec:para} and Table \ref{table:params}).

The results for $\tau_f$ = 1.0 and 4.0 Myr are shown in Figure \ref{fig-formation-rate}, where the radius of planetesimals ranges from 1 km to 500 km with the power-law index of $\alpha$ = 2.8.
When $\tau_f$ = 1 Myr, $\mathscr{F}_{\rm ratio} (T_{\rm th} \geq 600 ^\circ{\rm C})$ gets larger than that for the constant rate (see Figures \ref{fig-size-range} (a) and \ref{fig-formation-rate} (a)).
This means that, when the formation rate of planetesimals rapidly decreases with time, the contribution from earlier formed planetesimals, which experience higher temperatures, becomes stronger.
For $\tau_f$ = 4 Myr, the results are close to those with the constant formation rate (see Figures \ref{fig-size-range} (a) and \ref{fig-formation-rate} (b)).

From these results, we conclude that there is the distinguishable difference between the results for the constant formation rate and the decreasing formation rates with short $\tau_f$.

\section{Discussions} \label{sec:disc}

\begin{figure}
\plotone{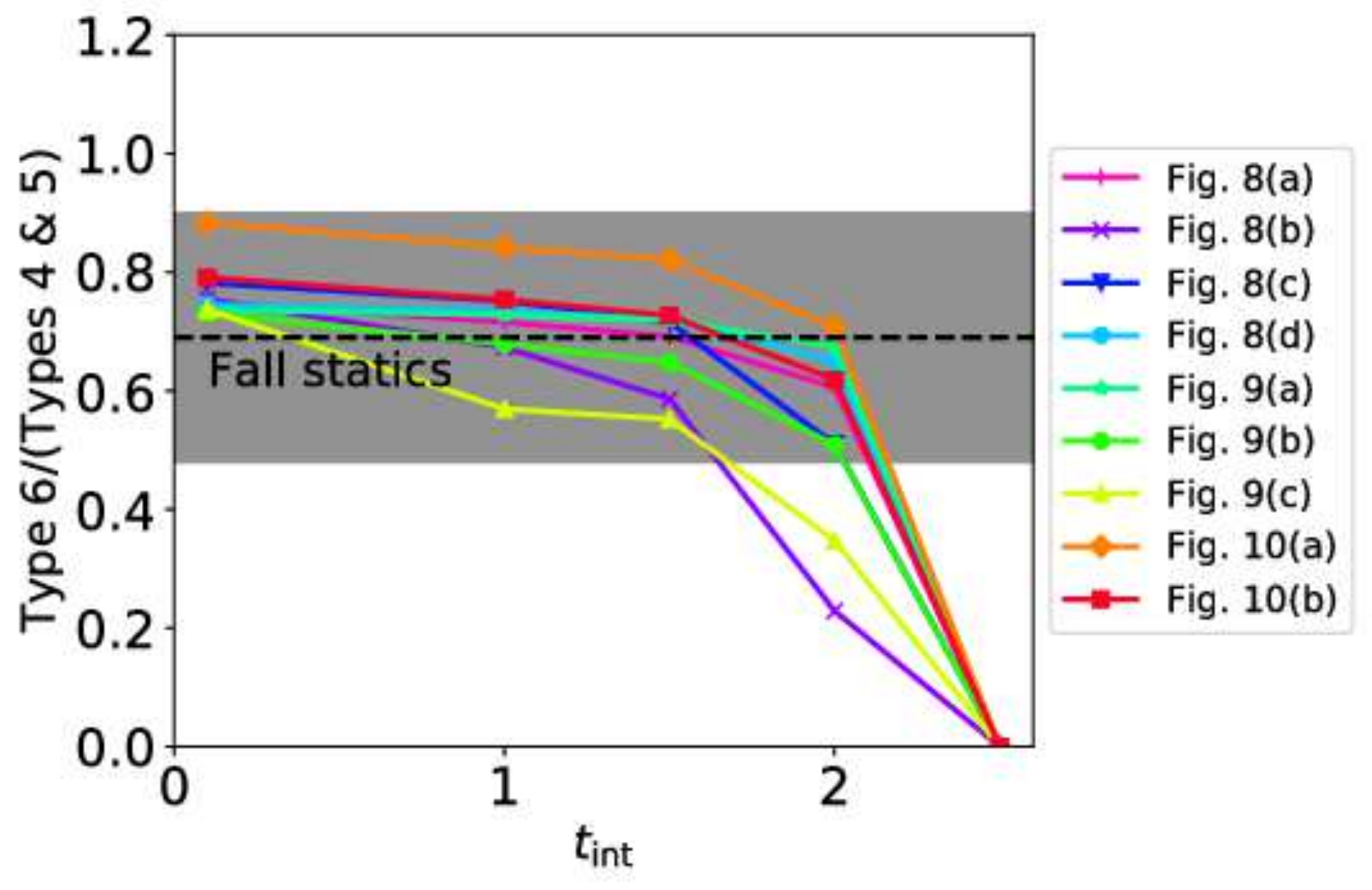}
\caption{Relative abundance ratios of type 6 to types 4 \& 5 ordinary chondrites. 
Each line comes from our results (see legends and corresponding figures). 
The fall statistics of ordinary chondrites (dashed line) with the uncertainty $\pm$ 30\% (shaded region) are also shown. 
\label{fig-metratio}}
\end{figure}

In this section, we compare our results with the abundance of each petrologic type of ordinary chondrites and discuss the formation and evolution histories of planetesimals.

The fall statistics of ordinary chondrites indicates that the number abundance of each petrologic types is
5.2\% for type 3, 16.5\% for type 4, 38.5\% for type 5, 37.9\% for type 6, and 1.9 \% for others \citep{gpm14}.
The petrologic types of chondrites are divided by chemical compositions and texture, as well as their peak metamorphic temperature.
Thus, the peak metamorphic temperature would not necessarily be adequate to separate each type rigorously; especially the estimated peak metamorphic temperatures are overlapped between types 4 and 5 \citep{sk05}. 
Nevertheless, in comparison to our results, we assign the following temperatures to each type:
100 - 600 $^\circ$C to type 3, 600 - 800  $^\circ$C to types 4 \& 5, and 800 - 1000 $^\circ$C to type 6 \citep{hrg06,sk05}.
In our simulations, we have computed the mass fraction of planetesimals ($\mathscr{F}_{\rm ratio} (T_{\rm th})$) for a variety of initial radii and formation times of planetesimals (see Equation (\ref{eq:ratio2})).
Although the fall statistics is given by the number of meteorites (by percentages), the statistics may represent the original (mass) abundance in their parent bodies, as a results of (thanks to) the huge number of meteorite samples.
Hence, we compare $\mathscr{F}_{\rm ratio} (T_{\rm th})$ with the fall statistics.

First, we find that, for all models considered in this study, $\mathscr{F}_{\rm ratio} (100 ^\circ {\rm C} \leq T_{\rm th} < 600 ^\circ {\rm C})$ is higher 
than $\mathscr{F}_{\rm ratio} (600 ^\circ {\rm C} \leq T_{\rm th} < 800 ^\circ {\rm C})$ and $\mathscr{F}_{\rm ratio} (800 ^\circ {\rm C} \leq T_{\rm th} \leq 1000 ^\circ {\rm C})$, meaning that type 3 ordinary chondrites are the most abundant in our models. 
This result is consistent with previous work \citep{mgg02}, but it is opposite to the fall statistics of type 3 chondrites, whose mass fraction is much lower than those of types 4 \& 5 and type 6 \citep{gpm14}.
We will discuss this point later.

Although our model cannot reproduce the fall statistics of type 3 ordinary chondrites, it is interesting to compare our results with the observed relative abundance of type 6 to types 4 \& 5 chondrites.
This is because these types of chondrites are dominant (90\% of all chondrites). 
Additionally, when thermal evolution of planetesimals is the main cause to generate types 4, 5, and 6 chondrites in their parent bodies, these chondrites should have been present in deep inside of the planetesimals (see Figure. \ref{fig:single}). 
Consequently, the relative abundance of type 6 to types 4 \& 5 chondrites could be a key to specifying formation times of planetesimals and their primordial sizes.
Figure \ref{fig-metratio} represents the mass ratios of type 6 to types 4 \& 5 chondrites obtained from our results as a function of the onset time of planetesimal formation (see legends).
In this plot, we consider some representative cases.
Note that $\mathscr{F}_{\rm ratio} (T_{\rm th} \geq 600 ^\circ {\rm C})$ becomes zero at $t_{\rm int} >$ 2.5 Myr, thus we cannot plot the result of $t_{\rm int} \geq$ 3.0 Myr.
Table \ref{tab-compare} summarizes our results for the case of $t_{\rm int}$ = 1.5 Myr, together with a set of parameters.

\begin{deluxetable*}{lllcl}
\tablecaption{Relative abundance ratios of type 6 to types 4 \& 5 ordinary chondrites calculated in this study. \label{tab-compare}}
\tablehead{
\colhead{Type 4} & \colhead{Type 5} & \colhead{Type 6} & \colhead{Type 6/Types 4\&5} & \colhead{Fall or Figure \& Parameters}
}
\startdata
0.18 & 0.41 & 0.41 & 0.69 & Fall statistics \\
\hline
\multicolumn{2}{c}{0.59} & 0.41 & 0.69 & Fig. \ref{fig-size-range} (a): $R_{\rm pl} = 1$ km - 500 km, $\alpha$ = 2.8 \\
\multicolumn{2}{c}{0.63} & 0.37 & 0.58 & Fig. \ref{fig-size-range} (b): $R_{\rm pl} = 1$ km - 50 km, $\alpha$ = 2.8  \\
\multicolumn{2}{c}{0.58} & 0.42 & 0.72 & Fig. \ref{fig-size-range} (c): $R_{\rm pl} = 60$ km - 90 km, $\alpha$ = 2.8  \\
\multicolumn{2}{c}{0.58} & 0.42 & 0.72 & Fig. \ref{fig-size-range} (d): $R_{\rm pl} = 100$ km - 500 km, $\alpha$ = 2.8  \\
\multicolumn{2}{c}{0.58} & 0.42 & 0.72 & Fig. \ref{fig-power-law} (a): $R_{\rm pl} = 1$ km - 500 km, $\alpha$ = 0.0  \\
\multicolumn{2}{c}{0.61} & 0.39 & 0.64 & Fig. \ref{fig-power-law} (b): $R_{\rm pl} = 1$ km - 500 km, $\alpha$ = 3.5  \\
\multicolumn{2}{c}{0.64} & 0.36 & 0.55 & Fig. \ref{fig-power-law} (c): $R_{\rm pl} = 1$ km - 500 km, $\alpha$ = 4.5  \\
\multicolumn{2}{c}{0.55} & 0.45 & 0.82 & Fig. \ref{fig-formation-rate} (a): $R_{\rm pl} = 1$ km - 500 km, $\alpha$ = 2.8, $\tau_f$ = 1.0 Myr \\
\multicolumn{2}{c}{0.58} & 0.42 & 0.72 & Fig. \ref{fig-formation-rate} (b): $R_{\rm pl} = 1$ km - 500 km, $\alpha$ = 2.8, $\tau_f$ = 4.0  Myr \\
\enddata
\tablecomments{Abundance ratio of each type is given for the case of $t_{\rm int}$ = 1.5 Myr and is normalized by the sum of the values for Types 4 to 6.}
\end{deluxetable*}

\begin{figure}
\plotone{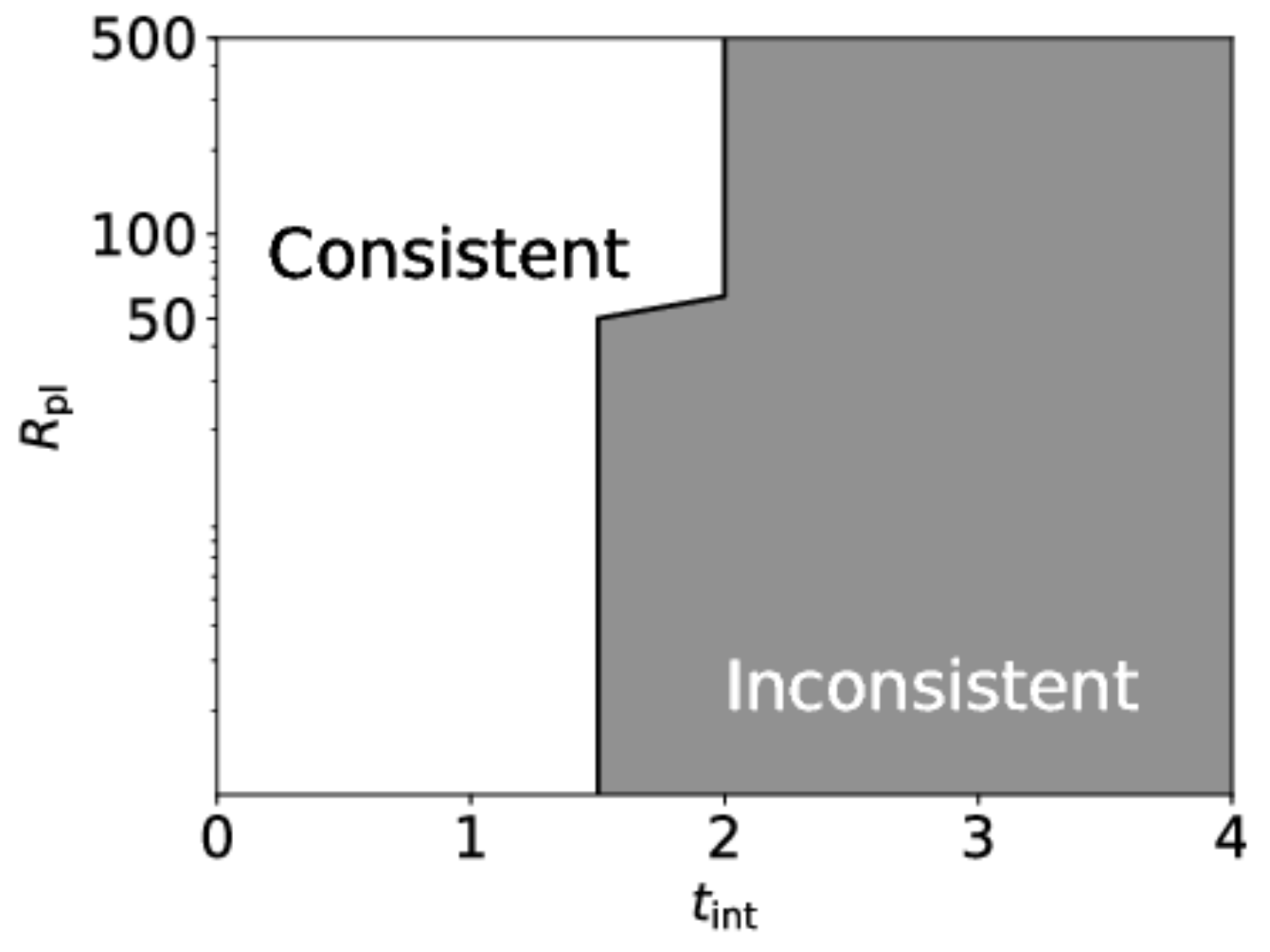}
\caption{Formation time of planetesimals and their sizes that are consistent with the ratio of type 6 to types 4 \& 5 ordinary chondrites obtained from the fall statistics.
We adopt $\alpha = 2.8$ and the constant formation rate. 
The parameter space that can reproduce the fall statistics is shown in the white region and the one that cannot is in the gray region. 
(see Figs. \ref{fig-size-range} in Table \ref{tab-compare}). \label{fig-last}}
\end{figure}

One of our important findings is that the parent bodies of ordinary chondrites must begin to form before $t \sim$ 2.0 Myr, to produce type 6 chondrites that undergo thermal metamorphism with the temperature of 800 $^\circ$C or higher. 
In other words, if formation of planetesimals would be initiated after $t \sim$ 2.0 Myr, no planetesimal could be heated high enough to contain type 6 chondrites, which contradicts the observational fact. 

It should be emphasized that the calculated mass ratios of type 6 to types 4 \& 5 are in agreement with the fall statistics within the uncertainty of 30\% for most of our models when $t_{\rm int} \leq$ 2 Myr.
The planetesimal formation time of 2.0 Myr is consistent with the estimation from our previous work for the parent bodies of the asteroid Itokawa \citep{wni14}.
However, adopting $t_{\rm int}$ = 2.0 Myr, the mass ratios of type 6 to types 4 \& 5 from the small size range ($R_{\rm pl}$ = 1 km - 50 km) of planetesimals become lower than the fall statistics (see Fig. \ref{fig-size-range} (b) in Figures \ref{fig-metratio} and Table \ref{tab-compare}).
The result with $\alpha$ = 4.5, wherein small planetesimals dominate the total mass, also underproduces the fall statistics (see Fig. \ref{fig-power-law} (c) in Figures \ref{fig-metratio} and Table \ref{tab-compare}). 
On the other hand, when planetesimal formation begins at $t \leq$ 1.5 Myr, these cases can successfully explain the fall statistics of type 6 ordinary chondrites.
Therefore, our results suggest two scenarios for the formation condition of planetesimals;
(1) planetesimals should have started to form at $t \leq$ 2.0 Myr after CAIs when their primordial sizes were larger than 60 km and/or
(2) planetesimals whose size distribution was weighted toward small radii of 1 km - 50 km should have started to form at $t \leq$ 1.5 Myr after CAI formation.
These conclusions are schematically depicted in Figure \ref{fig-last}.

\begin{deluxetable*}{ccccl}
\tablecaption{Relative abundance ratios in type 3 ordinary chondrites calculated in this study. \label{tab:type3}}
\tablehead{
\colhead{Types 3.0 - 3.2} & \colhead{Types 3.3 - 3.6} & \colhead{Types 3.7 - 3.9} & \colhead{Types 3.3-3.6/Types 3.7-3.9} & \colhead{Fall or Figure \& Parameters}
}
\startdata
0.13 & 0.40 & 0.47 & 0.84 & Fall statistics \\
\hline
0.83 & 0.09 & 0.08 & 1.16 & Fig. \ref{fig-size-range} (a): $R_{\rm pl} = 1$ km - 500 km, $\alpha$ = 2.8 \\
0.88 & 0.07 & 0.05 & 1.36 & Fig. \ref{fig-size-range} (b): $R_{\rm pl} = 1$ km - 50 km, $\alpha$ = 2.8  \\
0.82 & 0.10 & 0.08 & 1.21 & Fig. \ref{fig-size-range} (c): $R_{\rm pl} = 60$ km - 90 km, $\alpha$ = 2.8  \\
0.81 & 0.10 & 0.09 & 1.10 & Fig. \ref{fig-size-range} (d): $R_{\rm pl} = 100$ km - 500 km, $\alpha$ = 2.8  \\
0.81 & 0.10 & 0.09 & 1.06 & Fig. \ref{fig-power-law} (a): $R_{\rm pl} = 1$ km - 500 km, $\alpha$ = 0.0  \\
0.88 & 0.07 & 0.05 & 1.29 & Fig. \ref{fig-power-law} (b): $R_{\rm pl} = 1$ km - 500 km, $\alpha$ = 3.5  \\
0.93 & 0.04 & 0.03 & 1.52 & Fig. \ref{fig-power-law} (c): $R_{\rm pl} = 1$ km - 500 km, $\alpha$ = 4.5  \\
0.76 & 0.12 & 0.12 & 1.07 & Fig. \ref{fig-formation-rate} (a): $R_{\rm pl} = 1$ km - 500 km, $\alpha$ = 2.8, $\tau_f$ = 1.0 Myr \\
0.81 & 0.10 & 0.09 & 1.14 & Fig. \ref{fig-formation-rate} (b): $R_{\rm pl} = 1$ km - 500 km, $\alpha$ = 2.8, $\tau_f$ = 4.0  Myr \\
\enddata
\tablecomments{Abundance ratio of each subtype is given for the case of $t_{\rm int}$ = 1.5 Myr and is normalized by the sum of the values for types 3.0 to 3.9.}
\end{deluxetable*}

Figure \ref{fig-metratio} also shows that the results for size distributions with $\alpha$ = 0 and 3.5 lead to 
a little worse agreement with the fall statistics than that for the fiducial case with $\alpha$ = 2.8 (see Fig. \ref{fig-size-range} (a), Fig. \ref{fig-power-law} (a), and (b) in Figure \ref{fig-metratio} and Table \ref{tab-compare}). 
However, the difference is not so large that we cannot impose any constraint on the primordial size distribution of planetesimals.
Meanwhile, our results can provide some implications to the formation rate of planetesimals (see Fig. \ref{fig-size-range} (a), Fig. \ref{fig-formation-rate} (a), and (b) in Figures \ref{fig-metratio} and Table \ref{tab-compare}).
When the formation rate of planetesimals decreases slowly ($\tau_f=$4.0 Myr), 
the mass ratios of type 6 to types 4 \& 5 can reproduce the fall statistics relatively well (see Fig. \ref{fig-formation-rate} (b) in Figure \ref{fig-metratio} and Table \ref{tab-compare}).
Even for the rapid ($\tau_f$=1.0 Myr) case whose result is slightly higher than the fall statistics, it is in a reasonable range (see Fig. \ref{fig-formation-rate} (a) in Figure \ref{fig-metratio} and Table \ref{tab-compare}). 
Although the preferable formation rate of planetesimals can be a constant one or a relatively long formation timescale ($\tau_f$ = 4.0 Myr), we can not rule out the rapid formation case ($\tau_f$ = 1.0 Myr).

Our results show that the mass ratios of type 6 to types 4 \& 5 are produced by the thermal evolution model of planetesimals.
However, the mass fraction of type 3 ordinary chondrites cannot be explained.
As shown in Figures \ref{fig-size-range} - \ref{fig-formation-rate}, even if planetesimals start to form as early as 0.1 Myr, about 70\% of the total mass of the planetesimals cannot experience temperatures higher than 600 $^\circ{\rm C}$. 
Therefore, to lower the abundance of type 3 chondrites than those of types 4 - 6 ones, planetesimals have to be heated more than the present calculations. 
However, we do not know other internal sources and effective mechanisms of heating the whole planetesimals (expect for the decay of short-lived radionuclides).

\begin{figure}
\plotone{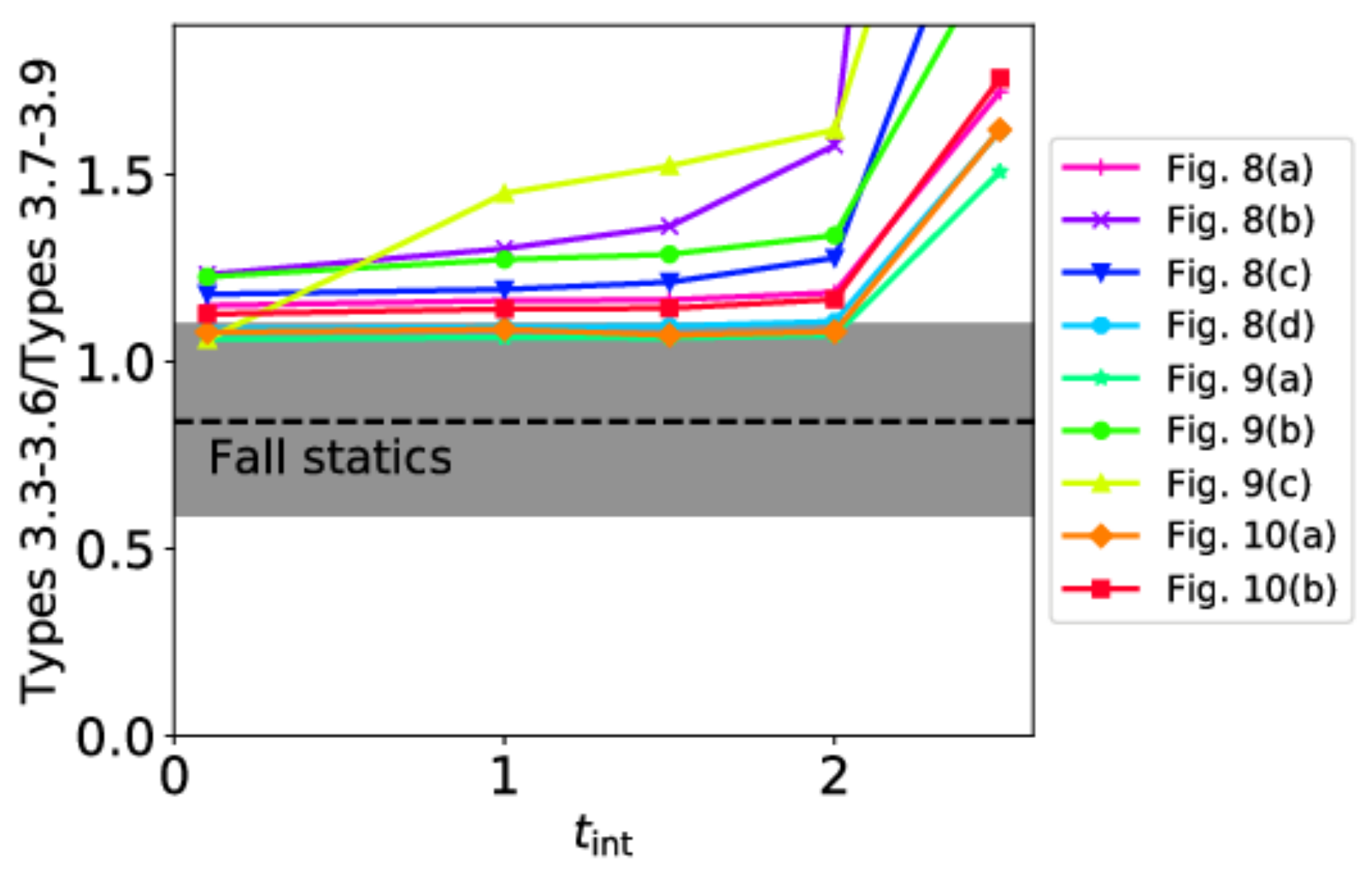}
\caption{Relative abundance ratios of types 3.3-3.6 to types 3.7-3.9 ordinary chondrites. 
Each line comes from our results (see legends and corresponding figures). 
The fall statistics of ordinary chondrites (dashed line) with the uncertainty $\pm$ 30\% (shaded region) are also shown.
\label{fig:type3}}
\end{figure}

Type 3 chondrites are divided into petrologic subtypes from 3.0 to 3.9 based on their characteristics of 
thermoluminescence sensitivity and chemical compositions of minerals, such as iron content of olivines \citep{sgmr80,sj90,sjr94,gb05,bbm06}. 
Although our model cannot reproduce the abundance of type 3 chondrites relative to types 4 \& 5 and type 6, 
it is still interesting to see what implications our results have for the fall statistics of type 3 subtypes. 
It is suggested that types 3.0 - 3.2 chondrites experienced much lower peak temperatures than the other type 3s 
and that type 3.9 would have experienced almost the same temperature as type 4 \citep[e.g.,][]{hrg06,cay08,vzb14}. 
It is very hard to distinguish the subtypes only by their peak metamorphic temperatures, but we here roughly divide type 3 subtypes into three subgroups, 
assigning peak metamorphic temperatures as follows: 100 - 400 $^\circ$C for types 3.0 - 3.2, 400 - 500 $^\circ$C for types 3.3 - 3.6, 
and 500 - 600 $^\circ$C for types 3.7 - 3.9. 
The relative fall statistics of type 3 subtypes are derived based on Meteoritical Bulletin Database (\url{http://www.lpi.usra.edu/meteor/metbull.php}): 
0.132, 0.398, and, 0.470 for types 3.0 - 3.2, 3.3 - 3.6, and 3.7 - 3.9, respectively. 
Note that we use only the samples of type 3 ordinary chondrites that are clearly classified as subtypes. 
We summarize the normalized fall statistics of type 3 subtypes and our results for the case of $t_{\rm int}$ = 1.5 Myr in Table \ref{tab:type3}. 
Our results show that the lower numbers of type 3 subtypes are more abundant than the higher numbers of them (i.e., types 3.0 - 3.2 $>$  3.3 - 3.6 $>$ 3.7 - 3.9), 
which is opposite to the trend appeared for the fall statistics (types 3.0 - 3.2 $<$ 3.3 - 3.6 $<$ 3.7 - 3.9). 
In particular, our models predict much higher abundances of types 3.0 - 3.2 ($>$ 0.8) than its fall statistics (0.13, see Table \ref{tab:type3}). 
Even if we take the ratio of types 3.3 - 3.6 to types 3.7 - 3.9, almost all of our results overestimate the observed values out of the uncertainty of 30 \% (Figure \ref{fig:type3}). 
This disagreement might partially stem from the facts that the fall statistics of type 3 subtypes used here are far from complete 
(only 1/3 of type 3 ordinary chondrites is sorted out into subtypes) and/or that we do not consider the heat transport through volatile materials, 
which would affect the abundances of the least metamorphosed ones (types 3.0 - 3.2).

One strong possibility to account for the small fall statistics of type 3 ordinary chondrites 
(as well as the tendency that less metamorphosed type 3 subtypes are less abundant) is impact events for planetesimals. 
In this study, we assume that planetesimals do not undergo any destructive processes once they formed. 
However, planetesimals which were not incorporated into (proto)planets would have suffered from destructive collisions with their surrounding planetesimals over the age of the solar system. 
In fact, destructive collisions are one of the important ingredients for reproducing the size distribution of the current main asteroid belt \citep{mbn09}.
When planetesimals experience destructive collisions, materials located in their surface regions are ejected.
Such materials can potentially serve as the parent bodies of type 3 chondrites due to a low level of metamorphism.
It nonetheless can be expected that they would not contribute to the (current) fall statistics very much.
This is because they would be probably accreted onto nearby (proto)planets and/or be gone beyond the solar system on a relatively short timescale.
Furthermore, the removal of surface materials via destructive collisions would increase the chance for the core region of planetesimals to be exposed to subsequent collisions.
Given that materials in the core region can become the parent bodies of types 4 - 6 chondrites, destructive collisions can eventually enhance the possibility that types 4 - 6 chondrites fall onto the Earth as meteorites.
It is interesting that the asteroid Itokawa would be mainly composed of types 5 - 6 chondrites, and its parent body is considered to have been originally larger than 20 km in radius before being catastrophically destroyed by a big impact \citep[e.g.,][]{nnt11,wni14}.
Thus, the destructive collisions that planetesimals have underwent during their long life may effectively decrease the amount of type 3 chondrites. 
In other words, the fall statistics of type 3 chondrites may hold the key information on the collision histories of planetesimals.

\section{Conclusions} \label{sec:conc}

In this paper, we systematically examine the thermal evolution of planetesimals with a wide range of initial sizes and formation times.
Our numerical results show that planetesimals can reach higher temperatures when they form at earlier times and/or have larger initial radii. 
There is a considerable difference in maximum temperatures between a small size range (from 1 km to 50 km) and a larger size one (from 60 km to 90 km and/or from 100 km to 500 km) (see Figure \ref{fig-size-range}). 

We also examine the effect of the formation rate of planetesimals.
We found that, in the case of the continuous formation of planetesimals, their mass fraction is significantly affected by the onset time of planetesimal formation (Figure \ref{fig-size-range}).
This trend is prominent when we focus on the planetesimal mass that experiences higher temperatures ($\geq$600$^\circ$C).
Furthermore, our results show that planetesimals should have form within 2.0 Myr after formation of Ca-Al-rich inclusions (CAIs), in order to produce type 6 ordinary chondrites, which would have experienced the peak metamorphic temperature of 800 $\sim$ 1000$^\circ$C.
We also found that the ratio of type 6 to types 4 \& 5 chondrites in fall statistics can be well explained by thermal evolution models of planetesimals.
This strongly suggests that these types of chondrites are produced mainly by thermal metamorphism inside the parent bodies.
Preferable scenarios to account for the fall statistics are that planetesimals with radii of $\geq 60$ km start to form around 2.0 Myr and/or that planetesimals of $\leq$ 50km start to form at $\leq$ 1.5 Myr after the CAIs formation.

On the other hand, our thermal evolution models of planetesimals cannot explain the fall statistics of type 3 ordinary chondrites; we predict that type 3 chondrites are unavoidably more abundant than any of types 4 to 6 chondrites.
This is against the fact that type 3 chondrites are the least abundant in the fall statistics.
The overabundance of type 3 chondrites could be resolved by taking into account subsequent destructive processes such as collisions between planetesimals.
In our future work, we will explore the effect of impact and subsequent destruction for better understanding the formation and evolution histories of planetesimals.

\acknowledgments
We thank the referee, Andrew, M. Davis for helpful comments and suggestions.
Numerical computations were carried out on the PC cluster at Center for Computational Astrophysics, National Astronomical Observatory of Japan.
T. N. has been supported in part by a JSPS Grant-in-Aid for Scientific Research (26400223).
The part of this research was carried out at the Jet Propulsion Laboratory, California Institute of Technology, under a contract with the National Aeronautics and Space Administration.
Y. H. is supported by JPL/Caltech.


\begin{thebibliography}{}
\bibitem[Amelin et al.(2010)]{aki10} Amelin, Y., Kaltenbach, A., Iizuka, T., et al.\ 2010, Earth and Planetary Science Letters, 300, 343
\bibitem[Blackburn et al.(2017)]{bac17} Blackburn, T., Alexander, C.~M.~O., Carlson, R., \& Elkins-Tanton, L.~T.\ 2017, \gca, 200, 201
\bibitem[Bland \& Travis(2017)]{bt17} Bland, P.~A., \& Travis, B.~J.\ 2017, Science Advances, 3, e1602514 
\bibitem[Bollard et al.(2015)]{bcb15} Bollard, J., Connelly, J.~N., \& Bizzarro, M.\ 2015, Meteoritics and Planetary Science, 50, 1197
\bibitem[Bonal et al.(2006)]{bbm06} Bonal, L., Quirico, E., Bourot-Denise, M., \& Montagnac, G.\ 2006, \gca, 70, 1849
\bibitem[Bottke et al.(2005a)]{bdn05} Bottke, W.~F., Durda, D.~D., Nesvorn{\'y}, D., et al.\ 2005, \icarus, 175, 111
\bibitem[Bottke et al.(2005b)]{bdn05b} Bottke, W.~F., Durda, D.~D., Nesvorn{\'y}, D., et al.\ 2005, \icarus, 179, 63 
\bibitem[Connelly et al.(2012)]{cbk12} Connelly, J.~N., Bizzarro, M., Krot, A.~N., et al.\ 2012, Science, 338, 651
\bibitem[Cody et al.(2008)]{cay08} Cody, G.~D., Alexander, C.~M.~O., Yabuta, H., et al.\ 2008, Earth and Planetary Science Letters, 272, 446
\bibitem[Cuzzi et al.(2001)]{chp01} Cuzzi, J.~N., Hogan, R.~C., Paque, J.~M., \& Dobrovolskis, A.~R.\ 2001, \apj, 546, 496
\bibitem[Davis \& McKeegan(2014)]{dm14} Davis, A.~M., \& McKeegan, K.~D.\ 2014, Meteorites and Cosmochemical Processes, 361
\bibitem[Davis et al.(2014)]{dac14} Davis, A.~M., Alexander, C.~M.~O.~'., Ciesla, F.~J., et al.\ 2014, Protostars and Planets VI, ed. H. Beuther et al. (Tucson, AZ: Univ. Arizona Press), 809
\bibitem[Dodd(1981)]{d81} Dodd, R.~T.\ 1981, Research supported by the U.S.~Air Force, State University of New York, NSF, and Max-Planck-Gesellschaft.~Cambridge, England and New York, Cambridge University Press, 1981.~377 p
\bibitem[Gail et al.(2014)]{gtb14} Gail, H.-P., Trieloff, M., Breuer, D., \& Spohn, T.\ 2014, Protostars and Planets VI, ed. H. Beuther et al. (Tucson, AZ: Univ. Arizona Press), 571 
\bibitem[Ghosh \& McSween(2000)]{gm00} Ghosh, A., \& McSween, H.~Y., Jr.\ 2000, Meteoritics and Planetary Science Supplement, 35, A59 
\bibitem[Grady et al.(2014)]{gpm14} Grady, M., Pratesi, G., \& Moggi Cecchi, V.\ 2014, Atlas of Meteorites, by Monica Grady , Giovanni Pratesi , Vanni Moggi Cecchi, Cambridge, UK: Cambridge University Press, 2014
\bibitem[Grimm \& McSween(1989)]{gm89} Grimm, R.~E., \& McSween, H.~Y., Jr.\ 1989, \icarus, 82, 244 
\bibitem[Grossman \& Brearley(2005)]{gb05} Grossman, J.~N., \& Brearley, A.~J.\ 2005, Meteoritics and Planetary Science, 40, 87  

\bibitem[Hasegawa et al.(2016a)]{hwmo16} Hasegawa, Y., Wakita, S., Matsumoto, Y., \& Oshino, S.\ 2016, \apj, 816, 8
\bibitem[Hasegawa et al.(2016b)]{htm16} Hasegawa, Y., Turner, N.~J., Masiero, J., et al.\ 2016, ApJL, 820, L12 
\bibitem[Henke et al.(2013)]{hgt13} Henke, S., Gail, H.-P., Trieloff, M., \& Schwarz, W.~H.\ 2013, \icarus, 226, 212
\bibitem[Huss et al.(2006)]{hrg06} Huss, G.~R., Rubin, A.~E., \& Grossman, J.~N.\ 2006, Meteorites and the Early Solar System II, ed. D. S. Lauretta and H. Y. McSween, Jr. (Tucson, AZ: Univ. Arizona Press), 567 
\bibitem[Ida \& Lin(2004)]{il04} Ida, S., \& Lin, D.~N.~C.\ 2004, \apj, 604, 388
\bibitem[Johansen et al.(2007)]{jom07} Johansen, A., Oishi, J.~S., Mac Low, M.-M., et al.\ 2007, \nat, 448, 1022
\bibitem[Johansen et al.(2014)]{jbmt14} Johansen, A., Blum, J., Tanaka, H., et al.\ 2014, in Protostars and Planets VI, ed. H. Beuther et al. (Tucson, AZ: Univ. Arizona Press), 547
\bibitem[Johansen et al.(2015)]{jmb15} Johansen, A., Mac Low, M.-M., Lacerda, P., \& Bizzarro, M.\ 2015, Science Advances, 1, 1500109 
\bibitem[Johnson et al.(2015)]{jmm15} Johnson, B.~C., Minton, D.~A., Melosh, H.~J., \& Zuber, M.~T.\ 2015, \nat, 517, 339
\bibitem[Keil(2000)]{k00} Keil, K.\ 2000, \planss, 48, 887
\bibitem[Kokubo \& Ida(1998)]{ki98} Kokubo, E., \& Ida, S.\ 1998, \icarus, 131, 171 
\bibitem[Kokubo \& Ida(2000)]{ki00} Kokubo, E., \& Ida, S.\ 2000, Icarus, 143, 15
\bibitem[Kominami et al.(2016)]{kdm16} Kominami, J.~D., Daisaka, H., Makino, J., \& Fujimoto, M.\ 2016, \apj, 819, 30 
\bibitem[Kretke \& Lin(2007)]{kl07} Kretke, K.~A., \& Lin, D.~N.~C.\ 2007, \apjl, 664, L55 
\bibitem[Krot et al.(2005)]{kkgs05} Krot, A.~N., Keil, K., Goodrich, C.~A., Scott, E.~R.~D., \& Weisberg, M.~K.\ 2005, Meteorites, Comets and Planets: Treatise on Geochemistry, Volume 1.~Edited by A.~M.~Davis.~Executive Editors: H.~D.~Holland and K.~K.~Turekian.~ISBN 0-08-044720-1.~Published by Elsevier B.~V., Amsterdam, The Netherlands, 2005, p.83, 83  
\bibitem[Levison et al.(2010)]{ltd10} Levison, H.~F., Thommes, E., \& Duncan, M.~J.\ 2010, \aj, 139, 1297 
\bibitem[Lichtenberg et al.(2016)]{lgg16} Lichtenberg, T., Golabek, G.~J., Gerya, T.~V., \& Meyer, M.~R.\ 2016, \icarus, 274, 350
\bibitem[Matsumoto et al.(2017)]{mohw17} Matsumoto, Y., Oshino, S., Hasegawa, Y., \& Wakita, S.\ 2017, \apj, 837, 103 
\bibitem[MacPherson et al.(1995)]{mdz95} MacPherson, G.~J., Davis, A.~M., \& Zinner, E.~K.\ 1995, Meteoritics, 30, 365
\bibitem[MacPherson(2005)]{m05} MacPherson, G.~J.\ 2005, Meteorites, Comets and Planets: Treatise on Geochemistry, Volume 1.~Edited by A.~M.~Davis.~Executive Editors: H.~D.~Holland and K.~K.~Turekian.~ISBN 0-08-044720-1.~Published by Elsevier B.~V., Amsterdam, The Netherlands, 2005, p.201, 201
\bibitem[McSween(1999)]{m99} McSween, H.~Y., Jr.\ 1999, Meteorites and their Parent Planets, by Harry Y.~McSween, Jr, pp.~322.~ISBN 0521587514.~Cambridge, UK: Cambridge University Press, February 1999., 322
\bibitem[McSween et al.(2002)]{mgg02} McSween, H.~Y., Jr., Ghosh, A., Grimm, R.~E., Wilson, L., \& Young, E.~D.\ 2002, Asteroids III, W. F. Bottke Jr., A. Cellino, P. Paolicchi, and R. P. Binzel (eds), University of Arizona Press, Tucson, p.559-571
\bibitem[Miller \& Fortney(2011)]{mf11} Miller, N., \& Fortney, J.~J.\ 2011, \apjl, 736, L29 
\bibitem[Miyamoto et al.(1982)]{mft82} Miyamoto, M., Fujii, N., \& Takeda, H.\ 1982, Lunar and Planetary Science Conference Proceedings, 12, 1145 
\bibitem[Monnereau et al.(2013)]{mtb13} Monnereau, M., Toplis, M.~J., Baratoux, D., \& Guignard, J.\ 2013, \gca, 119, 302
\bibitem[Morbidelli et al.(2009)]{mbn09} Morbidelli, A., Bottke, W.~F., Nesvorn{\'y}, D., \& Levison, H.~F.\ 2009, Icarus, 204, 558 
\bibitem[Mordasini et al.(2009)]{mab09} Mordasini, C., Alibert, Y., \& Benz, W.\ 2009, \aap, 501, 1139
\bibitem[Mordasini et al.(2016)]{mbm16} Mordasini, C., van Boekel, R., Molli{\`e}re, P., Henning, T., \& Benneke, B.\ 2016, \apj, 832, 41 
\bibitem[Morishima et al.(2010)]{msm10} Morishima, R., Stadel, J., \& Moore, B.\ 2010, Icarus, 207, 517 
\bibitem[Nakamura et al.(2011)]{nnt11} Nakamura, T., Noguchi, T., Tanaka, M., et al.\ 2011, Science, 333, 1113

\bibitem[Opeil et al.(2010)]{ocb10} Opeil, C.~P., Consolmagno, G.~J., \& Britt, D.~T.\ 2010, Icarus, 208, 449 
\bibitem[Pollack et al.(1996)]{phb96} Pollack, J.~B., Hubickyj, O., Bodenheimer, P., et al.\ 1996, \icarus, 124, 62 
\bibitem[Raymond et al.(2014)]{rkm14} Raymond, S.~N., Kokubo, E., Morbidelli, A., Morishima, R., \& Walsh, K.~J.\ 2014, Protostars and Planets VI, ed. H. Beuther et al. (Tucson, AZ: Univ. Arizona Press), 595 
\bibitem[Ricard et al.(2017)]{rba17} Ricard, Y., Bercovici, D., \& Albar{\`e}de, F.\ 2017, \icarus, 285, 103 
\bibitem[Saumon \& Guillot(2004)]{sg04} Saumon, D., \& Guillot, T.\ 2004, \apj, 609, 1170
\bibitem[Sears et al.(1980)]{sgmr80} Sears, D.~W., Grossman, J.~N., Melcher, C.~L., Ross, L.~M., \& Mills, A.~A.\ 1980, \nat, 287, 791
\bibitem[Scott \& Jones(1990)]{sj90} Scott, E.~R.~D., \& Jones, R.~H.\ 1990, \gca, 54, 2485 
\bibitem[Scott et al.(1994)]{sjr94} Scott, E.~R.~D., Jones, R.~H., \& Rubin, A.~E.\ 1994, \gca, 58, 1203 
\bibitem[Scott \& Krot(2005)]{sk05} Scott, E.~R.~D., \& Krot, A.~N.\ 2005, Meteorites, Comets and Planets: Treatise on Geochemistry, Volume 1.~Edited by A.~M.~Davis.~Executive Editors: H.~D.~Holland and K.~K.~Turekian.~ISBN 0-08-044720-1.~Published by Elsevier B.~V., Amsterdam, The Netherlands, 2005, p.143, 143 
\bibitem[Simon et al.(2016)]{say16} Simon, J.~B., Armitage, P.~J., Li, R., \& Youdin, A.~N.\ 2016, \apj, 822, 55 
\bibitem[Travis \& Schubert(2005)]{ts05} Travis, B.~J., \& Schubert, G.\ 2005, Earth and Planetary Science Letters, 240, 234
\bibitem[Tsirvoulis et al.(2018)]{tmd18} Tsirvoulis, G., Morbidelli, A., Delbo, M., \& Tsiganis, K.\ 2018, \icarus, 304, 14

\bibitem[Van Schmus \& Wood(1967)]{vw67} Van Schmus, W.~R., \& Wood, J.~A.\ 1967, \gca, 31, 747
\bibitem[Vernazza et al.(2014)]{vzb14} Vernazza, P., Zanda, B., Binzel, R.~P., et al.\ 2014, \apj, 791, 120
\bibitem[Vernazza et al.(2015)]{vzn15} Vernazza, P., Zanda, B., Nakamura, T., Scott, E.~R.~D., \& Russell, S.\ 2015, Asteroids IV, ed. P. Michel et al. (Tucson, AZ: Univ. Arizona Press) 617 
\bibitem[Wakita et al.(2014)]{wni14} Wakita, S., Nakamura, T., Ikeda, T., \& Yurimoto, H.\ 2014, Meteoritics and Planetary Science, 49, 228 
\bibitem[Wakita et al.(2017a)]{wmoh17} Wakita, S., Matsumoto, Y., Oshino, S., \& Hasegawa, Y.\ 2017, \apj, 834, 125
\bibitem[Wakita et al.(2017b)]{wnh17} Wakita, S., Nozawa, T., \& Hasegawa, Y.\ 2017, \apj, 836, 106  
\bibitem[Wetherill \& Stewart(1989)]{ws89} Wetherill, G.~W., \& Stewart, G.~R.\ 1989, \icarus, 77, 330 
\bibitem[Weisberg et al.(2006)]{wmk06} Weisberg, M.~K., McCoy, T.~J., \& Krot, A.~N.\ 2006, Meteorites and the Early Solar System II, ed. D. S. Lauretta and H. Y. McSween, Jr. (Tucson, AZ: Univ. Arizona Press), 19
\bibitem[Wlotzka(2005)]{w05} Wlotzka, F.\ 2005, Meteoritics and Planetary Science, 40, 1673
\bibitem[Yomogida \& Matsui(1983)]{ym83} Yomogida, K., \& Matsui, T.\ 1983, Meteoritics, 18, 430
\bibitem[Yoshikawa et al.(2015)]{ykf15} Yoshikawa, M., Kawaguchi, J., Fujiwara, A., \& Tsuchiyama, A.\ 2015, Asteroids IV, ed. P. Michel et al. (Tucson, AZ: Univ. Arizona Press), 397
\bibitem[Zheng et al.(2017)]{zlk17} Zheng, X., Lin, D.~N.~C., \& Kouwenhoven, M.~B.~N.\ 2017, \apj, 836, 207 
\end{thebibliography}
\end{document}